\documentclass{iopart}
\usepackage{graphicx}
\usepackage{bm}
\usepackage{amssymb,bbm,epsfig,cite}
\usepackage{iopams}

\newcommand{\assign}{:=}

\newcommand{\mathd}{\mathrm{d}}

\newcommand{\tmmathbf}[1]{\ensuremath{\boldsymbol{#1}}}
\newcommand{\tmtextbf}[1]{{\bfseries{#1}}}
\newcommand{\tmstrong}[1]{{\bfseries{#1}}}
\newcommand{\tmtextit}[1]{{\itshape{#1}}}

\newcommand{\tmem}[1]{{\em #1\/}}

\newcommand{\text}[1]{\mbox{\rm #1}}

\newcommand{\tmop}[1]{{\rm #1}}
\newcommand{\bignone}{}

\begin{document}
{\vspace*{-3\baselineskip}\hspace*{\fill}\sf published in: J.~Phys.~A:~Math.~Theor. 43, 015303 (2010) }

\jl{1}

\title{Pointer basis induced by collisional decoherence}

\author{Marc Busse$^1$ and Klaus Hornberger$^{1,2}$}
\address{$^1$Arnold Sommerfeld Center for Theoretical Physics, \\
Ludwig-Maximilians-Universit{\"a}t M{\"u}nchen, \\Theresienstra{\ss}e 37, 
80333 Munich, Germany
}
\address{$^2$Max Planck Institute for the Physics of Complex Systems, \\
N{\"o}thnitzer Stra{\ss}e 38, 
01187 Dresden, Germany
}

\date{\today}

\begin{abstract}
We study the emergence and dynamics of pointer states in the motion of a quantum test particle affected by collisional decoherence. 
These environmentally distinguished states are shown to be exponentially localized solitonic wave functions 
which evolve according to the classical equations of motion. We explain their formation using the orthogonal unraveling of the master equation, 
and we demonstrate that the statistical weights of the 
arising mixture are given by projections of the initial state onto the pointer basis.
\end{abstract}

\pacs{03.65.Yz, 47.45.Ab, 05.40.Jc}


\section{Introduction}

The influence of environmental degrees of freedom has been identified as the
key concept in explaining the classical behavior of macroscopic
systems in a quantum framework
{\cite{Joos1985a,Schlosshauer2007a,Zurek2003a}}. According to this notion a
preferred set of localized system states -- called the pointer basis
{\cite{zurek1981a}} -- is induced in the course of the interaction of the
system with its surrounding. Most characteristically, any initial
superposition of these pointer states gets rapidly mixed, while the only
states who retain their purity for a long time are the pointer states themselves.
While the basic ideas behind the decoherence process seems to be settled, it
still remains an open problem to understand the emergence, the dynamics, and
the main properties of the pointer states for microscopic realistic
environments.

Several strategies have been proposed so far for determining the pointer basis
given the environmental coupling. In {\cite{zurek1993a}} the suggestion was
made to sort all pure states in the Hilbert space according to their linear
entropy production rate. The pointer states are then identified with the
states having minimal loss of purity. Similar results are obtained by the
approach of {\cite{Diosi2000a,gisin1995a,strunz2002a}} which is based on a
time evolution equation whose solitonic solutions are identified with the
pointer states. So far, this concept has been applied to the damped harmonic
oscillator {\cite{gisin1995a,strunz2002a}} and to a free quantum particle
coupled linearly to a bath of harmonic oscillators
{\cite{Diosi2000a,strunz2002a}}. There, the solitonic solutions of the
corresponding nonlinear equation are coherent states and Gaussian wave
packets, respectively. Moreover, the decoherence to Gaussian pointer states
was proved to be generic for linear coupling models {\cite{eisert2004b}}.

In this paper, we extend the analysis from linear models to a non-perturbative
treatment of the interaction. We focus on the model of collisional
decoherence, which provides a realistic description of the decoherence process
generated by an ideal gas environment. Notably, experiments with interfering
fullerene molecules displayed a reduction of interference visibility in
agreement with this model {\cite{hornberger2003a,Hackermuller2003b,hornberger2004a,Vacchini2004a,Vacchini2005a}}. We derive
the corresponding pointer states which are shown to form an overcomplete,
exponentially localized set of basis states, and we prove decoherence to these
states using the orthogonal unraveling of the master equation
{\cite{Diosi1986a,Rigo1996a}}. This stochastic process on the one hand
provides the statistical weights of the pointer basis, and on the other hand
represents an efficient way of solving master equations which exhibit pointer
states. Moreover, it explains the emergence of classical Hamiltonian dynamics.

The main result was already announced in {\cite{Busse2009a}}. Here, we
provide a more detailed explanation of the proofs and derivations. As an
extension of {\cite{Busse2009a}}, we prove the decoherence dynamics for a more
general situation where the localization rate of collisional decoherence is in
a non-saturated regime, and we utilize the relative entropy in order to
illustrate the emergence of the statistical weights.

While the above results are derived within the framework of decoherence
theory, they can also be applied to dynamic reduction models which propose a
modification of the Schr\"odinger equation by means of nonlinear and
stochastic terms. In fact, the observational consequences of the
Ghirardi-Rimini-Weber (GRW) spontaneous localization model
{\cite{Ghirardi1986a,Bassi2003a}} are equivalent to the ones of collisional
decoherence, since they are described by the same master equation
{\cite{Vacchini2007b}}. The present work therefore applies to the GRW model.
In particular, it provides the corresponding pointer basis.

The structure of the article is as follows. In Sect.~2, we briefly review the
notion of pointer states, and we summarize the method for determining the
pointer states discussed in {\cite{gisin1995a,Diosi2000a,strunz2002a}}. In
order to motivate the approach, we consider a two level system and a dephasing
process. This method is then applied to collisional decoherence in Sect.~3,
which provides a set of solitonic states to be regarded as `candidate' pointer
states. We show in Sect.~4 that these solitons form an overcomplete basis of
exponentially localized states, and we give an expression for their spatial
extension. Moreover, we demonstrate that the `candidate' pointer states move on
classical phase space trajectories if they are sufficiently localized. In
Sect.~5, we briefly review the method of quantum trajectories, focusing in
particular on the orthogonal unraveling of the master equation. This
stochastic process is then applied to collisional decoherence which allows us
to show that the `candidate' states are indeed pointer states in the sense of
the definition given in Sect.~2. Furthermore, we use the orthogonal
unraveling to show that the statistical weights of the pointer states are
given by the overlap with the initial state. We present our conclusion in
Section 6. \ \ \ \ \ \ \ \ \ \ \

\section{The pointer basis{\tmem{{\tmem{{\tmem{{\tmem{}}}}}}}}}

\subsection{Definition of pointer states}

To motivate the definition of pointer states, let us consider the quantum
dynamics of the damped harmonic oscillator. Its evolution can be described by
a master equation in Lindblad form defined by the standard Hamiltonian
$\mathsf{H} = \hbar w \mathsf{a}^{\dag} \mathsf{a}$, and a single Lindblad
operator $\mathsf{L} = \mathsf{a}$, with associated rate $\gamma$
{\cite{Breuer2007b}}. We take as initial state a superposition of two
quasi-orthogonal coherent states
\begin{eqnarray}
  | \psi_0 \rangle & = & c_1 | \alpha_0 \rangle + c_2 | \beta_0 \rangle \,,
  \hspace{2em} \tmop{with} \,\, \left| \alpha_0 - \beta_0 \right|^2
  \gg 1 \,,  \label{eq:supercoherentstates}
\end{eqnarray}
which satisfy $\mathsf{a} | \alpha \rangle = \alpha | \alpha \rangle$, with
$\alpha \in \mathbbm{C}$. It is then easy to show that for times larger than
the decoherence time $t_{\tmop{dec}} = 2 \left| \alpha_0 - \beta_0 \right|^{-
2} / \gamma$ the solution of the master equation is well approximated by
{\cite{Hornberger2009a,Breuer2007b}}
\begin{eqnarray}
  \rho_t & \simeq & \left| c_1 \right|^2 | \alpha_t \rangle \langle \alpha_t |
  + \left| c_2 \right|^2 | \beta_t \rangle \langle \beta_t | \,, \hspace{2em}
  \tmop{if} \,\, t \gg t_{\tmop{dec}} \,, 
  \label{eq:mixturecoherentstates}
\end{eqnarray}
with $\alpha_t = \alpha_0 \, \exp \left( - iwt - \gamma t / 2 \right)$. Thus,
any coherent state remains pure during the damped time evolution, while any
superposition of distinct coherent states decays into a mixture (with a decay
rate $\gamma_{\tmop{dec}} = 1 / t_{\tmop{dec}}^{} \gg \gamma$) whose
statistical weights are determined by the initial overlaps $\left| \langle
\alpha_0 | \psi_0 \rangle \right|^2$ and $\left| \langle \beta_0 | \psi_0
\rangle \right|^2$. Due to this property, the coherent states are to be
identified with the pointer states of the damped harmonic oscillator.

The above observation serves as the starting point for the following
definition of the pointer states for an open quantum system evolving according
to a Lindblad master equation $\partial_t\rho =\mathcal{L} \rho$. One says that
the system exhibits a pointer basis if its dynamics involves a separation of
time scales, characterized by a fast decoherence time $t_{\tmop{dec}}$, such
that for any time much larger than $t_{\tmop{dec}}$, the evolved state is well
approximated by a mixture of uniquely defined pure states $\mathsf{P}_{\alpha}
= | \pi_{\alpha} \rangle \langle \pi_{\alpha} |$ which are independent of the
initial state $\rho_0$,
\begin{eqnarray}
  e^{\mathcal{L}t} \rho_0 & \simeq & _{} \int \mathd \alpha \, \,
  \bignone_{\bignone} \tmop{Prob} \left( \alpha | \rho_0 \right)
  \mathsf{P}_{\alpha} (t), \hspace{2em} \tmop{if} \,\, \text{$t \gg t_{\tmop{dec}}$} \,, 
  \label{eq:decoherence}
\end{eqnarray}
with $\tmop{Prob} \left( \alpha | \rho_0 \right) \geqslant 0, \int \mathd
\alpha \, \tmop{Prob} \left( \alpha | \rho_0 \right) = 1 \bignone$. Following
the above example, we further demand that for initial states $\rho_0$, which
are superpositions of mutually orthogonal pointer states $\mathsf{P}_{\beta}$,
the probability distribution $\tmop{Prob} \left( \alpha | \rho_0 \right) =
\sum_{\beta} w_{\beta} \delta \left( \alpha - \beta \right) \bignone$ is given
by the initial projections
\begin{eqnarray}
  w_{\beta} & = & \tmop{Tr} \left( \rho_0 \mathsf{P}_{\beta} \left( 0 \right)
  \right) \, .  \label{eq:bornrule}
\end{eqnarray}
The pointer states $\mathsf{P}_{\alpha}$ initially form an (overcomplete)
basis, and they may evolve in time, though slowly compared to
$t_{\tmop{dec}}$.

The name {\tmem{pointer state}} was coined in {\cite{zurek1981a}} due to its
relevance for the physical description of a measurement apparatus. A
measurement device which probes an observable $\mathsf{A}$ is constructed such
that macroscopically distinct positions of the pointer or indicator are
obtained for the different eigenstates of $\mathsf{A}$. For a quantum system
initially prepared in an eigenstate of $\mathsf{A}$ the read-out will display
the corresponding eigenvalue with certainty provided these pointer states
remain pure during the time evolution. On the other hand, if the quantum
system is prepared in a superposition of eigenstates of $\mathsf{A}$, we
expect the pointer not to end up in a superposition of different read-out
states, but rather to be at a definite position, though probabilistically,
with probabilities given by the overlap (\ref{eq:bornrule}). \

We emphasize that the importance of pointer states goes beyond the physics of
measurement devices and the quantum-to-classical transition since they are
also a practical tool for the solution of master equations. Knowing the
pointer states $\mathsf{P}_{\alpha}$, their time evolution
$\mathsf{P}_{\alpha} \left( t \right)$, and their probabilities $\tmop{Prob}
\left( \alpha | \rho_0 \right)$, one can immediately specify the solution of
the master equation for any initial state and times greater than the
decoherence time. Since the decoherence time is generically much shorter than
the system and dissipation time scales of the pointer state motion, this
allows one to capture a large part of the system evolution without solving the
master equation. \

\subsection{Pointer states of pure dephasing \ }

{\normalsize{}}A practical way to obtain the pointer states
$\mathsf{P}_{\alpha}$, for a given environmental coupling, was discussed in
{\cite{Diosi2000a,gisin1995a,strunz2002a}}. We will illustrate this method by
means of a two level system, suspect to a dephasing environment. The
corresponding master equation in interaction picture,
\begin{eqnarray}
  \partial_t \rho & \text{$= \,$} & \gamma \left( \sigma_z \, \rho \, \sigma_z
  - \rho \right) \,,  \label{eq:dephasing}
\end{eqnarray}
is characterized by the Lindblad jump operator $\mathsf{L =} \sigma_z
\sqrt{\gamma}$ $\left( \gamma > 0 \right)$. In Bloch representation, $\rho_t =
1 / 2 \left( \mathsf{I} + \tmmathbf{a}_{} \left( t \right) \cdot
\tmmathbf{\sigma} \right)$ with Bloch vector $\tmmathbf{a} \tmmathbf{_{}}_{}
\left( t \right) = \tmop{Tr} \left( \tmmathbf{\sigma} \rho \left( t \right)
\right)$ and Pauli matrices $\tmmathbf{\sigma} = \left( \sigma_x, \sigma_y,
\sigma_z \right)$, the solution reads
\begin{eqnarray}
  \tmmathbf{a} \tmmathbf{_{}}_{} \left( t \right) & = & \left( e^{- 2 \gamma
  t} a_x \left( 0 \right), e^{- 2 \gamma t} a_y \left( 0 \right), a_z \left( 0
  \right) \right) \, .  \label{eq:decayblochvector}
\end{eqnarray}
Thus, the Bloch sphere $\left\{ \tmmathbf{a}: \left| \tmmathbf{a} \right| = 1
\right\}$, which represents the set of all pure states, is projected onto the
z-axis in the course of the dephasing process. This implies that the decohered
state $\tmmathbf{a} \left( \infty \right) = \left( 0, 0, a_z \left( 0 \right)
\right)$ is a mixture of the eigenstates of $\sigma_z$ (denoted by
$\mathsf{P}_{\downarrow} = | \downarrow \rangle \langle \downarrow |$ and
$\mathsf{P}_{\uparrow} = | \uparrow \rangle \langle \uparrow |$ respectively),
\begin{eqnarray}
  \rho_{\infty} & = & _{} \, \bignone_{\bignone} \tmop{Tr} \left[
  \mathsf{P_{\uparrow}}_{} \rho_0 \right] \mathsf{P_{\uparrow}} + \tmop{Tr}
  \left[ \mathsf{P_{\downarrow}}_{} \rho_0 \right] \mathsf{P_{\downarrow} \,}
  .  \label{eq:decoherencedephasing}
\end{eqnarray}
The comparison with (\ref{eq:decoherence}) shows that the north and the south
pole of the Bloch sphere (corresponding to $\mathsf{P}_{\downarrow}$ and
$\mathsf{P}_{\uparrow}$) form the pointer basis of the pure dephasing process.

Since the solution of the master equation will not be at hand in general, one
requires a method which yields the pointer states without the knowledge of
$\rho_{t \,}$. To motivate this, we notice that the north pole of the Bloch
sphere is the asymptotic end point of the trajectory illustrated by the thick
line in Fig.~1. This trajectory is generated by an equation of motion with the
following properties. ({\tmem{i}}) It is nonlinear because it must distinguish
pointer states from their superpositions. ({\tmem{ii}}) It preserves the
purity of pure initial states, i.e. an initial state which lies on the Bloch
sphere remains on the surface. ({\tmem{iii}}) The generated trajectory follows
the exact solution of the master equation as close as possible. In order to
find such an equation of motion, it is suggestive to minimize the distance of
the initial increments, i.e.
\begin{eqnarray}
  \min_{\partial_t \tmmathbf{a}} |\mathcal{L} \left( \tmmathbf{a} \right) -
  \partial_t \tmmathbf{a} |^2 \,, \hspace{2em} \tmop{with} \; \tmmathbf{a}
  \cdot \partial_t \tmmathbf{a}= 0 \, . &  &  \label{eq:minimum}
\end{eqnarray}
Here, $\mathcal{L}$ denotes the generator of the master equation
(\ref{eq:dephasing}) in Bloch representation and $\partial_t \tmmathbf{a}$ is
subject to the condition that the generated trajectory remains on the surface
of the Bloch sphere. The solution of the above optimization problem reads, in
spherical coordinates, \
\begin{eqnarray}
  \text{$\left( \dot{r}, \dot{\varphi}, \dot{\theta} \right)$} & = & \left( 0,
  0, - \gamma \sin \left( 2 \theta \right) \right) \, . 
  \label{eq:kugelkoordinaten}
\end{eqnarray}
Since the sine is positive for $\theta \in \left( 0, \pi / 2 \right)$, the
solutions of these equations tend asymptotically towards a pointer state of
the system, see Fig.~1. The equator of the Bloch sphere forms a set of
unstable fixed points of (\ref{eq:kugelkoordinaten}).

\begin{figure}[tb]
  \begin{flushright}
    \begin{center}
      \resizebox{5cm}{!}{\includegraphics{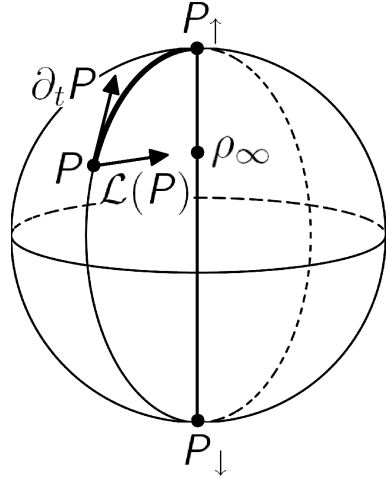}}
    \end{center}
  \end{flushright}
  \caption{Bloch representation of a two level system subject to pure
  dephasing, as described by (\ref{eq:dephasing}). As $t \rightarrow \infty$,
  the initial state $\rho_0 = \mathsf{P}$ is projected onto the z-axis,
  implying that the poles ($\mathsf{P}_{\uparrow}$ and
  $\mathsf{P}_{\downarrow} \,$) are the pointer states. The thick line
  indicates a trajectory within the set of pure states which connects the
  initial state to a nearby pointer state (the north pole). The equation of
  motion for this trajectory has minimal distance from $\mathcal{L} \left(
  \mathsf{P} \right)$ among all equations which generate pure state
  trajectories.  \ }
\end{figure} \ \ \ \ \ \ \ \ \ \ \ \ \ \ \ \ \ \ \ \ \ \ \ \ \ \ \

\subsection{Nonlinear equation for pointer states}

Let us now generalize the above argument to general Markovian master
equations. Replacing the Euclidean norm by the Hilbert-Schmidt norm $\| \cdot
\|_{\rm{HS}}$ in the space of operators, the generalization of
(\ref{eq:minimum}) to higher dimensional systems reads
\begin{eqnarray}
  \min_{\mathsf{} \partial_t \mathsf{P}} ||\mathcal{L} \left( \mathsf{P}
  \tmmathbf{} \right) - \mathsf{} \partial_t \mathsf{P} ||_{\rm{HS}}^2 \,, &
  &  \label{eq:minimum2}
\end{eqnarray}
where the minimization is with respect to all evolution equations $\mathsf{}
\partial_t \mathsf{P} = f \left( \mathsf{P} \right)$ which propagate
$\mathsf{P}$ within the set of pure states, such that $\mathsf{P}_t^2 =
\mathsf{P}_t \,$. It can be shown that the general structure of an equation
that evolves state vectors $| \psi \rangle$ preserving their normalization has
the structure $\partial_t | \psi \rangle = \left( \mathsf{A}_{\psi} - \langle
\psi | \mathsf{A}_{\psi} | \psi \rangle + \mathsf{B}_{\psi} \right) | \psi
\rangle$, with $\psi$-dependent, hermitian and anti-hermitian mappings
$\mathsf{A}_{\psi} = \mathsf{A}_{\psi}^{^{\dag}}$ and $\mathsf{B}_{\psi} = -
\mathsf{B}_{\psi}^{^{\dag}} \,$. This implies that the equation of motion for
the projector $\mathsf{P} = | \psi \rangle \langle \psi |$ must be of the form
$\partial_t \mathsf{P} = \left[ \mathsf{P}, \left[ \mathsf{P},
\mathsf{X}_{\mathsf{P}} \right] \right] \,,$where $\mathsf{X}_{\mathsf{P}}
\assign \mathsf{A}_{\mathsf{P}} + \left[ \mathsf{B}_{\mathsf{P}}, \mathsf{P}
\right]$. Using this form, the optimization problem (\ref{eq:minimum2})
reduces to $\min_{\mathsf{X}} \| \mathcal{L} \left( \mathsf{P} \right) -
\left[ \mathsf{P}, \left[ \mathsf{P}, \mathsf{X}_{} \right] \right]
\|^{}_{\rm{HS}} \,_{}$. As shown in {\cite{gisin1995a}} the solution is determined by the
generator of the master equation, $\mathsf{X}_{\min} = \mathcal{L} \left(
\mathsf{P} \right)$. Hence, the generalization of (\ref{eq:kugelkoordinaten})
reads as {\cite{Diosi2000a,gisin1995a,strunz2002a}}
\begin{eqnarray}
  \partial_t \mathsf{P} & = & \left[ \mathsf{P}, \left[ \mathsf{P},
  \mathcal{L} \left( \mathsf{P} \right)]] \, .  \label{eq:NG} \right. \right.
\end{eqnarray}
Motivated by the example in Sect.~2.2, one conjectures that the asymptotic
solutions of (\ref{eq:NG}) provide the pointer states in more complex systems
as well.

It will be important below that Eq.~(\ref{eq:NG}) is known also in another context: it
corresponds to the deterministic part of the orthogonal unraveling of the
master equation. As we will demonstrate in Section 5, one can use this
specific unraveling to prove for a specific model that the asymptotic
solutions of (\ref{eq:NG}) indeed provide the pointer states.

\section{Pointer states of collisional decoherence}

\subsection{Collisional decoherence}

In order to assess the nonlinear equation (\ref{eq:NG}) in the context of a
nontrivial environmental coupling we now apply it to the model of collisional
decoherence {\cite{Gallis1990a,Hornberger2003b}}. The latter describes the
motion of a quantum test particle in an ideal gas environment and it accounts
for the quantum effects of the scattering dynamics in a non-perturbative
fashion. The corresponding master equation has Lindblad form,
\begin{eqnarray}
  \partial_t \rho & = & - \frac{i}{\hbar} \left[ \mathsf{H}, \rho \right] +
  \int \mathd q \bignone \left( \mathsf{L}_q \rho \mathsf{L}^{^{\dag}}_q -
  \frac{1}{2} \mathsf{L}^{^{\dag}}_q \mathsf{L}_q \rho - \frac{1}{2} \rho
  \mathsf{L}^{^{\dag}}_q \mathsf{L}_q \right) \,,  \label{eq:lindblad}
\end{eqnarray}
where the jump operators are proportional to momentum kick operators,
$\mathsf{L}_q = \sqrt{\gamma G \left( q \right)} e^{iq \mathsf{x}}$ (with position operator $\mathsf{x}$). The
continuous label $q$ has the meaning of a momentum transfer experienced by the
test particle with $G \left( q \right)\geq 0$ the corresponding distribution,
$\int \mathd q\, \bignone G \left( q \right) = 1$; $\gamma$ is the
collision rate of the gas environment. The 1d equation of motion thus reads
\begin{eqnarray}
  \partial_t \rho & = & \frac{1}{i \hbar} \left[ \mathsf{H}, \rho \right] +
  \gamma \int_{- \infty}^{\infty} \mathd q \bignone \, G \left( q \right)
  e^{iq \mathsf{x} / \hbar} \rho e^{- iq \mathsf{x} / \hbar} - \gamma \rho \,
  .  \label{eq:CD}
\end{eqnarray}
It leads to a localization in position space, i.e. to a loss of spatial
coherence, as can be seen by switching to the interaction picture,
$\tilde{\rho} = e^{i \mathsf{H} t / \hbar} \rho e^{- i \mathsf{H} t / \hbar}$,
and the position representation,
\begin{eqnarray}
  \partial_t \langle x| \tilde{\rho} |x' \rangle & = & - F \left( x - x'
  \right) \langle x| \tilde{\rho} |x' \rangle \, .  \label{eq:decaycoherences}
\end{eqnarray}
The decay rate of the spatial coherences is thus characterized by a
{\tmem{localization rate}} $F \left( s \right)\geq 0$ which is related to the
momentum transfer distribution $G \left( q \right) \,$ by
\begin{eqnarray}
  F \left( s \right) & = & \gamma \left( 1 - \int_{- \infty}^{\infty} \mathd q
  \bignone \, G \left( q \right) e^{i \tmop{qs} / \hbar} \right) \, . 
  \label{eq:localization}
\end{eqnarray}
Since the Fourier transform of the distribution $G \left( q \right)$ tends to
zero for large distances $s$, the localization rate saturates for large $s$ at
the maximum value given by the collision rate, $F \left( s \rightarrow
\infty \right) = \gamma \,$, which can be interpreted as the limit where one
collision is sufficient to reveal the particles `which path' information. This
behavior is in sharp contrast to linear models, where the localization rate
grows quadratically, and thus approaches infinity in the limit of a large 
separation $s$.

\begin{figure}[tb]

  \begin{center}
    \resizebox{12cm}{!}{\includegraphics{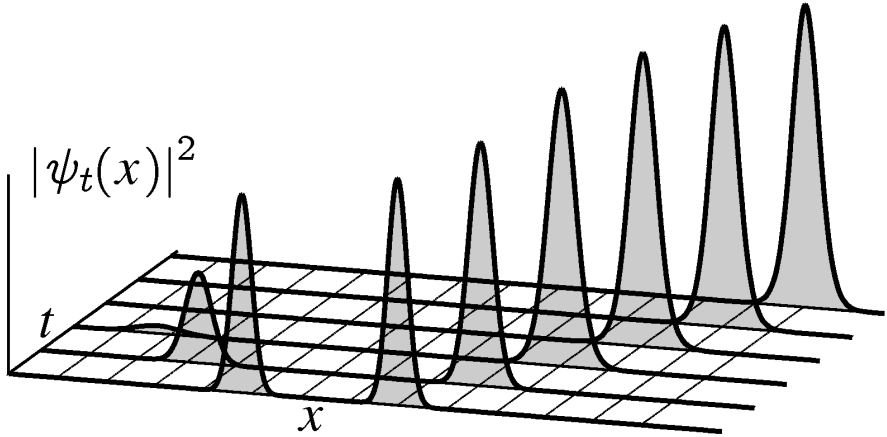}}
  \end{center}
  \caption{Formation of pointer states: an initial superposition of
  counter-propagating, localized states is evolved numerically according to
  the nonlinear equation (\ref{eq:dimensionless}). It forms into a solitonic
  solution which moves with a fixed shape and constant velocity. These
  solitons are interpreted as the pointer states of collisional decoherence. }
\end{figure}

\subsection{Determining the pointer states of collisional decoherence}

In order to apply the nonlinear equation (\ref{eq:NG}) to collisional
decoherence (\ref{eq:CD}) of a free particle, that is $\mathsf{H} =
\mathsf{p}^2 / 2 m \,$, we rewrite the projector equation (\ref{eq:NG}) in
vector representation and choose for $\mathcal{L}$ the Lindblad form
(\ref{eq:lindblad}), which gives {\cite{gisin1995a,strunz2002a}}
\begin{eqnarray}
 \fl \partial_t | \psi \rangle & = & \frac{1}{i \hbar} \left( \mathsf{H} -
  \langle \mathsf{H} \rangle_{\psi} \right) \mathsf{} | \psi \rangle
  \nonumber\\\fl 
  &  & + \int_{- \infty}^{\infty} \mathd q \bignone \, \left( \langle
  \mathsf{L}^{\dag}_q \rangle_{\psi} \left( \mathsf{L_q} - \langle
  \mathsf{L}_q \rangle_{\psi} \right) - \frac{1}{2} \left(
  \mathsf{\mathsf{L}^{\dag}_q}_{} ^{} \mathsf{L_{}}_q - \langle
  \mathsf{\mathsf{L}^{\dag}_q}_{}  \mathsf{L_{}}_k \rangle_{\psi} \right)
  \right) \bignone | \psi \rangle \, .  \label{eq:nonlinear}
\end{eqnarray}
The expectation value $\langle \mathsf{H} \rangle_{\psi}$ is disregarded in
the following, since it contributes only an additional phase. Now, we choose
the jump operator of collisional decoherence, $\mathsf{L}_q = \sqrt{\gamma G
\left( q \right)} e^{iq \mathsf{x}}$, and switch to position representation,
which yields
\begin{eqnarray}
  \fl  \partial_t \psi_t \left( x \right) & = & - \frac{\hbar^{}}{2 mi}
  \partial_x^2 \psi_t \left( x \right) + \psi_t \left( x \right) \Lambda
  \left[ \left| \psi_t \right|^2 \right] \left( x \right) \,, 
  \label{eq:NGCD}\\ \fl 
  \Lambda \left[ \left| \psi_t \right|^2 \right] \left( x \right) & = & \gamma
  \left( | \psi_t |^2 \ast \hat{G} \left( x \right) - \int^{\infty}_{- \infty}
  \mathd y| \psi_t |^2 \left( y \right)  \left( | \psi_t |^2 \ast \hat{G}
  \right) \left( y \right) \right) \,.  \label{eq:lambda}
\end{eqnarray}
Here, $g \ast h (x) \equiv \int^{\infty}_{- \infty} \mathd y g \left( y
\right) h \left( x - y \right)$ denotes the convolution and $\hat{G} \left( x
\right)$ is the Fourier transform of $G \left( q \right) \,$, i.e. $\hat{G}
\left( x \right) \equiv \int_{- \infty}^{\infty} \mathd q \bignone \, G \left(
q \right) \exp \left( iqx / \hbar \right) \,$.

The two summands in (\ref{eq:NGCD}) have counteractive effects on the
temporal evolution of the wave function: the coherent term leads to its
dispersion, whereas the second, incoherent summand tends to localize the
solution. In order to explain this localization, we note that the second
summand in (\ref{eq:lambda}) is independent of $x$. This implies that the
centered parts of the wave function, where the convolution $| \psi_t |^2 \ast
\hat{G} \left( x \right)$ exceeds the constant term in (\ref{eq:lambda}), get
amplified, i.e. $\partial_t \psi_t > 0$, whereas the tails of the wave
function get damped, i.e. $\partial_t \psi_t < 0$. As a consequence of these
competing effects, solutions of (\ref{eq:NGCD}) evolve towards solitonic
states where both effects are in equilibrium, such that the state moves with
fixed shape and constant velocity. As discussed above, these solitons are
candidates for the pointer states of collisional decoherence.

Assuming the momentum transfer distribution $G \left( q \right)$ to be a
centered Gaussian with variance $\sigma_G^2 \,$, we can rewrite
(\ref{eq:NGCD}) in dimensionless form.
\begin{eqnarray}
  \partial_{\tau} \varphi_{\tau} \left( y \right) & = & - \frac{\kappa}{2 i}
  \partial_y^2 \varphi_{\tau} \left( y \right) + \varphi_{\tau} \left( y
  \right) \int_{- \infty}^{\infty} \mathd y' \bignone | \bignone
  \varphi_{\tau} \left( y' \right) |^2 \nonumber\\
  &  & \times \left( e^{- \left( y - y' \right)^2 / 2} - \int^{\infty}_{-
  \infty} \mathd y'' \bignone | \bignone \varphi_{\tau} \left( y'' \right) |^2
  e^{- \left( y' - y'' \right)^2 / 2} \right)  \label{eq:dimensionless}
\end{eqnarray}
Here we use the dimensionless variables $y \equiv \sigma_G x / \hbar \,$ and
$\tau \equiv \gamma t \,$ to define the dimensionless wave function
$\varphi_{\tau} \left( y \right) \equiv \sqrt{\hbar / \sigma_G} \psi_{\tau /
\gamma} \left( \hbar y / \sigma_G \right) \,$. Notably,
Eq.~(\ref{eq:dimensionless}) depends only on the single dimensionless
parameter $\kappa \equiv \sigma_G^2 / \left( m \hbar \gamma \right) \,$.

Figure~2 shows a numerical solution of (\ref{eq:dimensionless}), where we choose
a superposition of two counter propagating localized states $\phi_{1, 2}$ as
the initial state, $\psi_0 \left( x \right) = c_1 \phi_1 \left( x \right) +
c_2 \phi_2 \left( x \right)$. As expected from the above discussion, the
(modulus of the) solution converges to a soliton. Moreover, we find that the
soliton inherits its initial position and momentum expectation value from that
localized component $\phi_i$ of the initial state which has the greatest
weight $c_i$, $\left| c_i \right| > \left| c_{j \neq i} \right|$. Similar
observations are found for various other initial states.

\section{Properties of the soliton basis}

We proceed to characterize the solitonic solutions of (\ref{eq:NGCD}). In
Sect.~4.1, the consequences of the conservation of probability on the phase of the
solitons are analyzed, allowing us to predict the asymptotic shape of the
solitons in Sect.~4.2. In Sect.~4.3 we estimate the spatial extension of the
solitons, followed by the proof that they form a basis of the Hilbert space in
Sect.~4.4. Finally, in Sect.~4.5, we discuss the dynamics of the solitonic
solutions in the presence of an external potential.

\subsection{Consequences of the continuity equation}

As demonstrated in the previous section, the nonlinear equation
(\ref{eq:NGCD}) exhibits solitonic solutions $\pi_t \left( x \right)$ in the
sense that the modulus of $\pi_t \left( x \right)$ moves with constants shape
and velocity, i.e.
\begin{eqnarray}
  \pi_t \left( x \right) & = & f \left( x - vt \right) e^{ig \left( x, t
  \right)} \,,  \label{eq:shape}
\end{eqnarray}
with $f \geqslant 0$ and $g$ real. In this section, we analyze the general
structure of the phase $g \left( x, t \right)$, which will be relevant
subsequently. The time derivative of a solution $\left| \psi_t \left( x
\right) \right|^2$ of (\ref{eq:NGCD}), yields the continuity equation for
$\psi_t \left( x \right)$,
\begin{eqnarray}
  \partial_t \text{$\left| \psi_t \left( x \right) \right|^2 \,$ } & = & -
  \frac{\hbar}{m} \partial_x \tmop{Im} \left( \psi_t^{\ast} \partial_x \psi_t
  \right) + 2 \text{$\left| \psi_t \left( x \right) \right|^2 \,$} \Lambda
  \left[ | \psi_t |^2 \right] \left( x \right)  \, .  \label{eq:continuity}
\end{eqnarray}
Plugging the solitonic solution (\ref{eq:shape}) into (\ref{eq:continuity}),
gives
\begin{eqnarray}
 \fl - 2 \Lambda \left[ f^2 \right] \left( x - vt \right) - v \partial_x \log f^2
  \left( x - vt \right) & = & - \frac{\hbar}{m} (\partial_x^2 g \left( x, t
  \right) \nonumber\\
  &  & \left. + \partial_x^{} g \left( x, t \right) \partial_x \log f^2
  \left( x - vt \right) \right) \, .  \label{eq:intermediate1}
\end{eqnarray}
Here we have used that $\Lambda \left[ f_t^2 \right] \left( x \right) =
\Lambda \left[ f^2 \right] \left( x - vt \right)$, which follows from $f_t
\left( x \right) = f \left( x - vt \right)$\,. The time dependence of the left
hand side of (\ref{eq:intermediate1}) corresponds to a spatial shift. Thus,
also the right hand side of (\ref{eq:intermediate1}) must exhibit such a
simple time dependence, which implies that
\begin{eqnarray}
  \text{$- v \partial_x r \left( x, t \right) \,$} & = & \partial_t r \left(
  x, t \right) \,,  \label{eq:righthandside}
\end{eqnarray}
where $r \left( x, t \right)$ denotes the right hand side of
(\ref{eq:intermediate1}). It follows that
\begin{eqnarray}
 \fl - v \partial_x^3 g \left( x, t \right) - v \partial_x^2 g \left( x, t
  \right) \partial_x \log f^2 \left( x - vt \right) & = & \partial_t
  \partial^2_x g \left( x, t \right) \nonumber\\ &  & + \partial_t \partial_x g \left( x, t
  \right) \partial_x \log f^2 \left( x - vt \right) \, . 
    \label{eq:langegleichung}
\end{eqnarray}
Since this equation must hold for all $x, v$ and $t$, we may assume that the
equality holds already for the summands, such that
\begin{eqnarray}
  - v \partial_x^2 g \left( x, t \right) & = & \partial_t \left[ \partial_x g
  \left( x, t \right) \right] \, .  \label{eq:ableitung1}
\end{eqnarray}
Therefore, the temporal and spatial dependence of the phase has the general
structure
\begin{eqnarray}
  g \left( x, t \right) & = & \phi \left( x - vt \right) + \chi \left( t
  \right) \,,  \label{eq:phase}
\end{eqnarray}
with unknown functions $\phi$ and $\chi$.

\subsection{Asymptotic form of the solitons}

To explore the tails of the solitonic states $\pi_t \left( x \right)$ let us
consider the form of (\ref{eq:NGCD}) for asymptotically large positions. It
reads
\begin{eqnarray}
  \partial_t \psi_t \left( x \right) & \sim & - \frac{\hbar^{}}{2 mi}
  \partial_x^2 \psi_t \left( x \right) - \gamma a_{\psi} \psi_t \left( x
  \right) \,, \hspace{2em} \tmop{for}\,\,  \left| x \right|
  \rightarrow \infty \,,  \label{eq:asymptotic}
\end{eqnarray}
with $a_{\psi} \equiv \int^{\infty}_{- \infty} \mathd y| \psi_t |^2 \left( y
\right)  \left( | \psi_t |^2 \ast \hat{G} \right) \left( y \right)$ a
$\psi$-dependent, positive constant. Inserting the solitonic form
(\ref{eq:shape}) into (\ref{eq:asymptotic}) yields
\begin{eqnarray}
 \fl  i \partial_t g \left( x, t \right) f \left( x - vt \right) & = & i
  \frac{\hbar}{2 m} \left[ \partial_x^2 f \left( x - vt \right) - f \left( x -
  vt \right) \left( \partial_x g \left( x, t \right) \right)^2 \right] + vf
  \left( x - vt \right) \nonumber  \\
  &  & - \frac{\hbar}{m} \left[ \partial_x f \left( x - vt \right) \partial_x
  g \left( x, t \right) + f \left( x - vt \right) \partial_x^2 g \left( x, t
  \right) \right]\nonumber  \\ & & - \gamma a_{\psi} f \left( x - vt \right) \, . \label{eq:long}
\end{eqnarray}
Using (\ref{eq:phase}), we find that both $\partial_x g \left( x, t \right)$
and $\partial_x^2 g \left( x, t \right)$ are only a function of $x_t = x -
vt$, and accordingly, that also the left hand side of (\ref{eq:long}) must be
a function of $x_t$. If follows that $\chi \left( t \right)$ is at most linear
in $t \,$ (that is $\chi \left( t \right) = \chi_1 t + \chi_0 \,$, with
unknown constants $\chi_0$ and $\chi_1$). Considering the real and imaginary
part of (\ref{eq:long}) separately, one obtains two coupled (second order)
differential equations \
\begin{eqnarray}
  v \partial_x f - \gamma a_{\psi} f & = & \frac{\hbar}{m} \left( \partial_x f
  \partial_x \phi + \frac{1}{2} f \partial_x^2 \phi \right) \,, 
  \label{eq:set1}\\
  \left( \chi_1 - v \partial_x \phi \right) f & = & \frac{\hbar}{2 m} \left(
  \partial_x^2 f - f \left[ \partial_x \phi \right]^2 \right) \,, 
  \label{eq:set2}
\end{eqnarray}
where $f \equiv f \left( x - vt \right)$ and $\phi \equiv \phi \left( x - vt
\right)$. This set of equations has two unique solutions
\begin{eqnarray}
  f \left( x \right) & = & e^{\pm k \left| x \right|} \,,  \label{eq:f}\\
  \phi \left( x \right) & = & \mp \tmop{sgn} \left( x \right) \frac{m}{\hbar}
  \left( v + \frac{\gamma a_{\psi}}{k} \right) x,  \label{eq:phiofx}
\end{eqnarray}
where the constant $k > 0$ depends on the boundary condition for
(\ref{eq:set1}) (which can be determined only by solving the full nonlinear
equation (\ref{eq:NGCD})). The solution with the positive exponent in
(\ref{eq:f}) is irrelevant, since it is not normalizable. Figure 3 confirms
that the tails of the numerically obtained solitonic solutions of
(\ref{eq:dimensionless}) are in agreement with the functional form
(\ref{eq:f}); they are straight lines in the semi-logarithmic plot. This shows
that, unlike in linear models {\cite{eisert2004a}} where the pointer states
are Gaussian, the pointer states of collisional decoherence are exponentially
localized.

\begin{figure}[tb]

  \begin{center}
    \resizebox{9cm}{!}{\includegraphics{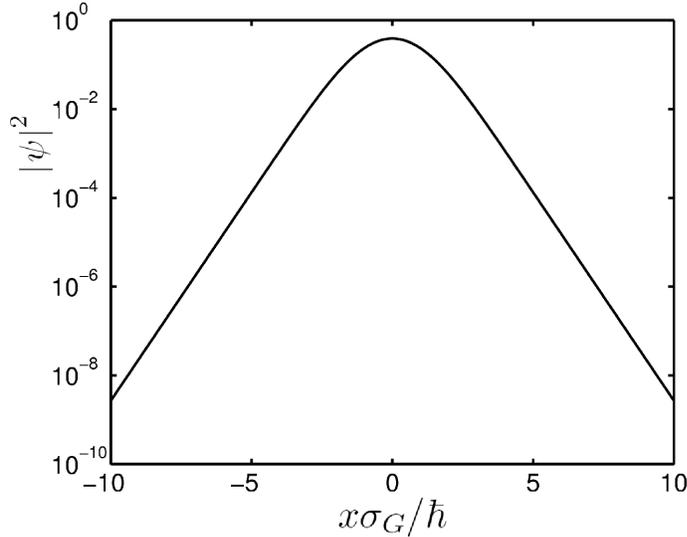}}
  \end{center}
  \caption{Semi-logarithmic plot of the numerical solitonic solution of
  (\ref{eq:dimensionless}). The graph clearly demonstrates that the pointer
  states have exponential tails.}
\end{figure}

\subsection{Size of the solitons}

An important characteristic of the pointer states is their spatial extension.
As explained in {\cite{Busse2009a}}, the latter can be related to the
experimentally accessible one-particle coherence length of a thermal gas. We
will determine the pointer width in this section, and apply the result later,
when studying the dynamics of pointer states in an external potential.

As a first step, consider the standard deviation $\tilde{\sigma}_{\pi}$ of
the numerically obtained dimensionless solitonic solution $\left| \tilde{\pi}
\left( y \right) \right|^2$ of (\ref{eq:dimensionless}) as a function the
dimensionless parameter $\kappa = \sigma_G^2 / \left( \gamma m \hbar \right)
\,$. As shown by the solid line in Fig.~4, the size $\tilde{\sigma}_{\pi}$
increases linearly with $\kappa \,$.

This observation can be reproduced by a simplified model which has the
practical advantage that it can be applied to more complex systems, such as 3D
gases with a microscopically realistic localization rate $F \,$
{\cite{Busse2009a}}. The idea of the model goes as follows: the ideal gas
environment consists of particles which collide with the system at a rate
$\gamma$. At each collision, the ambient particles gain position information,
such that the wave function gets spatially localized to a length scale
$\ell_{\tmop{loc}} \,$ determined by the localization rate $F$, see
(\ref{eq:decaycoherences}). After the scattering event, the particle disperses
freely, until it gets localized again by a subsequent collision. The
pointer width $\sigma_{\pi}$ is then obtained by averaging the time-dependent
width of the wave function over the waiting time distribution of a Poisson
process.

More specifically, we assume that the length scale $\ell_{\tmop{loc}}$ is
characterized by the free parameter
\begin{eqnarray}
  \text{$a_{\tmop{loc}}'$} & = & \frac{F \left( \ell_{\tmop{loc}} \right)}{F
  \left( \infty \right)}  \, .  \label{eq:characterizationlengthscale}
\end{eqnarray}
Using (\ref{eq:localization}) and taking the momentum transfer distribution $G
\left( q \right)$ to be a centered Gaussian with variance $\sigma_{^{} G^{}
\,}^2$, we obtain
\begin{eqnarray}
  \ell_{\tmop{loc}} & = & \frac{a_{\tmop{loc}} \, \hbar}{\sigma_G}_{} \:, 
  \label{eq:lloc}
\end{eqnarray}
with $a^2_{\tmop{loc}} = - 2 \log \left( 1 - a_{\tmop{loc}}' \right)^{}$. The
free dispersion after the collision yields the time dependent size
\begin{eqnarray}
  \sigma^2_{\pi} \left( t \right) & = & \ell_{\tmop{loc}}^2 + \left(
  \frac{\hbar t}{2 m \ell_{\tmop{loc}}} \right)^2 \, .  \label{eq:dispersion}
\end{eqnarray}
Finally, the average over the waiting time distribution $\tmop{Prob} \left( t
\right) = \gamma e^{- \gamma t}$ gives
\begin{eqnarray}
  \sigma_{\pi} & \equiv & \int_0^{\infty} \mathd \tau \tmop{Prob} \left( \tau
  \right) \bignone \frac{1}{\tau} \int_0^{\tau} \mathd t \sigma_{\pi} \left( t
  \right) \bignone \nonumber\\
  & \simeq & a_{\tmop{loc}} \frac{\hbar}{\sigma_G} + \frac{\sigma_G}{4
  a_{\tmop{loc}} m \gamma} \,,  \label{eq:size}
\end{eqnarray}
where we use a linearization of $\sigma_{\pi} \left( t \right)$ in the second
line. The dimensionless version of (\ref{eq:size}) reads
\begin{eqnarray}
  \tilde{\sigma}_{\pi} & \equiv & \frac{\sigma_{\pi} \sigma_G}{\hbar}
  \nonumber\\
  & = & a_{\tmop{loc}} + \frac{1}{4 a_{\tmop{loc}}} \kappa \, . 
  \label{eq:sizedimensionless}
\end{eqnarray}
The dashed line in Figure 4 shows that the form of this equation agrees with
the numerical solution of (\ref{eq:dimensionless}); the fit yields a value of $a_{\tmop{loc}} \simeq 0.4$ for the parameter characterizing the localization length scale.

\begin{figure}[tb]

  \begin{center}
    \resizebox{11cm}{!}{\includegraphics{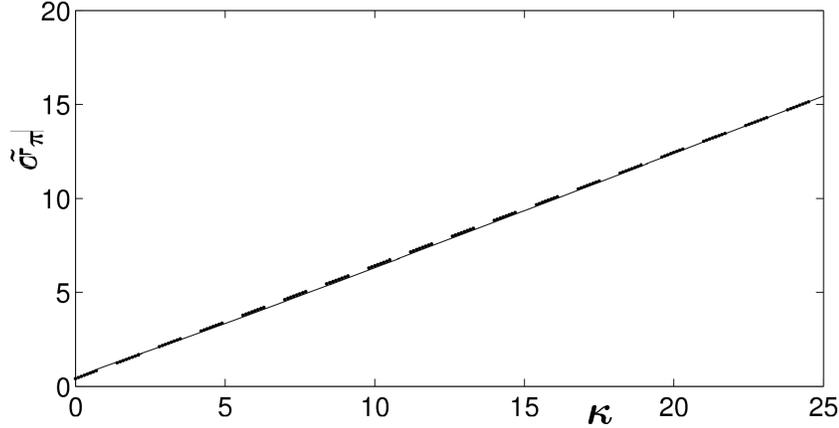}}
  \end{center}
  \caption{Spatial extension of the solitonic solution of
  (\ref{eq:dimensionless}) as a function of the dimensionless parameter
  $\kappa = \sigma_G^2 / \left( \gamma m \hbar \right)$. The solid line
  represents the numerical solution of (\ref{eq:dimensionless}). The result of
  the localization model (\ref{eq:sizedimensionless}) with parameter
  $a_{\tmop{loc}} = 0.4$ is given by the dashed line. }
\end{figure}

\subsection{Completeness of the soliton basis}

Our next aim is to show that the solitonic solutions of (\ref{eq:NGCD}), which
are interpreted below as the pointer states of collisional decoherence, form
an overcomplete basis. For this purpose, we first present a general method to
construct a whole manifold of solutions of (\ref{eq:NG}) given a specific one.
It relies on the symmetry properties of the corresponding master equation.
Since collisional decoherence exhibits Galilean (i.e. translation and boost)
invariance, it is then easy to show that the pointer states of this model form
an overcomplete basis. 

Suppose there is a family of unitary operators $\mathsf{U_t}$, satisfying
\begin{eqnarray}
  \hspace{1em} \mathsf{\mathsf{U}_t} \, \mathcal{D} \left( \rho \right) \,
  \mathsf{U}_t^{^{\dag}} & = & \mathcal{D} \left( \mathsf{\mathsf{U}_t} \,
  \rho \, \mathsf{U}_t^{^{\dag}} \right),  \label{eq:Condition1}\\
  \hspace{4em} \partial_t \, \mathsf{\mathsf{U}_t} & = & \frac{1}{i \hbar}
  \left[ \mathsf{H}, \mathsf{\mathsf{U}_t} \right],  \label{eq:Condition2}
\end{eqnarray}
where we denote by $\mathcal{D}$ the incoherent part of the master equation,
$\mathcal{L} \left( \rho \right) \equiv \left[ \mathsf{\mathsf{H}}, \rho
\right] / \left( i \hbar \right) +\mathcal{D} \left( \rho \right)$. Then,
given a solution $P_t$ of the nonlinear equation (\ref{eq:NG}), also
$\mathsf{\mathsf{\mathsf{\mathsf{U}_t} \mathsf{P_t}} \,
\mathsf{U}_t^{^{\dag}}}$ constitutes a solution of (\ref{eq:NG}).

This can be verified easily:
\begin{eqnarray}
 \fl  \left[ \text{$\mathsf{U} \mathsf{P} \mathsf{U}^{\dag}$}, \left[ \mathsf{U}
  \mathsf{P} \mathsf{U}^{\dag}, \mathcal{L} \left( \text{$\mathsf{U}
  \mathsf{P} \mathsf{U}^{\dag}$} \right) \right] \right]  & = & \frac{1}{i
  \hbar} \left( \mathsf{H} \mathsf{U} \mathsf{P} \mathsf{U}^{\dag} -
  \mathsf{U} \mathsf{P} \mathsf{U}^{\dag} \mathsf{H} \right) + \mathsf{U}
  \left[ \mathsf{P}, \left[ \mathsf{P}, \mathcal{D} \left( \mathsf{P} \right) 
  \right] \right] \mathsf{U}^{\dag}  \nonumber\\
  & = & { \frac{1}{i \hbar} ( \mathsf{H} \mathsf{U} \mathsf{P}
  \mathsf{U}^{\dag} - \mathsf{U} \mathsf{P} \mathsf{U}^{\dag} \mathsf{H} +
  \mathsf{U} \mathsf{H} \mathsf{P} \mathsf{U}^{\dag} - \mathsf{U} \mathsf{H}
  \mathsf{P} \mathsf{U}^{\dag} + \mathsf{U} \mathsf{P} \mathsf{H}
  \mathsf{U}^{\dag}} \nonumber\\
  &  & - \mathsf{U} \mathsf{P} \mathsf{H} \mathsf{U}^{\dag}) + \mathsf{U}
  \left[ \mathsf{P}, \left[ \mathsf{P}, \mathcal{D} \left( \mathsf{P} \right) 
  \right] \right] \mathsf{U}^{\dag}  \nonumber\\
  & = & \frac{1}{i \hbar} \left[ \mathsf{H}, \mathsf{U} \right] \mathsf{P}_{}
  \mathsf{U}^{\dag} + \mathsf{U} \left[ \mathsf{P}_{}, \left[ \mathsf{P}_{},
  \frac{1}{i \hbar} \left[ \mathsf{H}, \mathsf{P} \right] +\mathcal{D} \left(
  \mathsf{P}_{} \right) \right] \right] \mathsf{U}^{\dag} \nonumber\\
  &  & - \frac{1}{i \hbar} \mathsf{U} \mathsf{P}_{} \left[ \mathsf{U}^{\dag},
  \mathsf{H} \right]  \nonumber\\
  & = & \partial_t \left( \text{$\mathsf{U} \mathsf{P} \mathsf{U}^{\dag}$}
  \right) \,,  \label{eq:UPU}
\end{eqnarray}
where we dropped the time argument for brevity. Here, the first equality makes
use of (\ref{eq:Condition1}) and the unitarity of $\mathsf{U}$, and the third
equality is due to the relation $\left[ \mathsf{H}, \mathsf{P} \right] =
\left[ \mathsf{P}, \left[ \mathsf{P}, \left[ \mathsf{H}, \mathsf{P} \right]
\right] \right]$. In the last line, we use (\ref{eq:NG}) and
(\ref{eq:Condition2}).

Let us now apply this to the Galilean invariance of master equation
(\ref{eq:CD}). We will see that the phase space translations $\mathsf{U}_t
\equiv \mathsf{T}_{s, u} = \exp \left( i \left( u_t \mathsf{x} - s_t
\mathsf{p} \right) \right)$ satisfy the symmetry conditions
(\ref{eq:Condition1}) and (\ref{eq:Condition2}) provided the time
dependence of $s_t$ and $u_t$ has the particular form
\begin{eqnarray}
  s_t & = & u_0 \, t / m + s_0 \,,  \label{eq:time1}\\
  u_t & = & \text{$u_0 \,$} \, .  \label{eq:time2}
\end{eqnarray}
The latter enact a phase space translation in accordance with the free
shearing motion.

Let us first verify condition (\ref{eq:Condition1}) using $\mathsf{T}_{s, u} f \left( \mathsf{x}
\right) \mathsf{T}_{s, u}^{\dag} = f \left( \mathsf{x} - s \right)$.
\begin{eqnarray}
 \fl  \mathsf{T}_{s, u} \, \mathcal{D} \left( \rho \right) \mathsf{T}_{s,
  u}^{\dag} & = & \gamma \int_{- \infty}^{\infty} \mathd q \bignone \, G
  \left( q \right) \mathsf{T}_{s, u} \, e^{iq \mathsf{x} / \hbar} \rho \, e^{-
  iq \mathsf{x} / \hbar} \mathsf{T}^{\dag}_{s, u} - \gamma \mathsf{T}_{s, u}
  \rho \mathsf{T}^{\dag}_{s, u} \nonumber\\
  & = & \gamma \int_{- \infty}^{\infty} \mathd q \bignone \, G \left( q
  \right) \, e^{iq \left( \mathsf{x} - s \right) / \hbar} \mathsf{T}_{s, u}
  \rho \mathsf{T}_{s, u}^{\dag} \, e^{- iq \left( \mathsf{x} - s \right) /
  \hbar} - \gamma \mathsf{T}_{s, u} \rho \mathsf{T}^{\dag}_{s, u} \nonumber\\
  & = & \, \mathcal{D} \left( \mathsf{T}_{s, u} \, \rho \mathsf{T}_{s,
  u}^{\dag} \right) \,  \label{eq:TPT}
\end{eqnarray}
In order to verify (\ref{eq:Condition2}), use the Campbell-Hausdorff formula
to rewrite the translation operator as
\begin{eqnarray}
  \mathsf{T}_{s, u} & = & \exp \left( i \frac{u_t \mathsf{x}}{\hbar} \right)
  \, \exp \left( - i \frac{s_t \mathsf{p}}{\hbar} \right) \, \exp \left( - i
  \frac{s_t u_t}{2 \hbar} \right) \, .  \label{eq:translationoperator}
\end{eqnarray}
The time derivative thus yields
\begin{eqnarray}
  \partial_t \mathsf{} \mathsf{T}_{s, u} & = & \frac{i}{\hbar} \left(
  \dot{u}_t \mathsf{T}_{s, u} - \dot{s}_t \mathsf{T}_{s, u} \mathsf{p} -
  \frac{1}{2} \left( \dot{u}_t s_t + \dot{s}_t u_t \right) \mathsf{T}_{s, u}
  \right) \nonumber\\
  & = & \frac{i}{\hbar} \left( - \frac{u_t}{m} \mathsf{T}_{s, u} \mathsf{p} -
  \frac{u_t^2}{2 m} \mathsf{T}_{s, u} \right) \nonumber\\
  & = & \frac{1}{i \hbar} \left[ \frac{\mathsf{p}^2}{2 m}, \mathsf{T}_{s, u}
  \right] \,,  \label{eq:timederivativeofT}
\end{eqnarray}
where the shearing transformation (\ref{eq:time1}) and (\ref{eq:time2}) is
required in the second line. This confirms (\ref{eq:Condition2}) for
$\mathsf{H} = \mathsf{p}^2 / 2 m \,$.

We conclude that the nonlinear equation (\ref{eq:NGCD}) exhibits a family of
solitonic solutions $\mathsf{P}_{\Gamma} = \mathsf{T}_{s, u} \mathsf{P}
\mathsf{T}_{s, u}^{^{\dag}}$, parameterized by the phase space coordinate
$\Gamma = \left( s_0, u_0 \right)$. In order to verify that this family forms
an overcomplete basis, let us consider a specific class of phase space
representations. According to {\cite{klauder2007a}}, any Hilbert-Schmidt
operator $\mathsf{A}$ can be represented as
\begin{eqnarray}
  \mathsf{A} & = & \int \mathd \Gamma A \left( \Gamma \right) \bignone
  \mathsf{T}_{s, u} \mathsf{\mathsf{Q}} \mathsf{T}_{s, u}^{^{\dag}} \,, 
  \label{eq:phasespaceintegral}
\end{eqnarray}
provided $\mathsf{Q}$ is a trace-class operator, i.e. $0 < \tmop{Tr} \left(
\sqrt{\mathsf{Q}^{\dag} \mathsf{Q}} \right) < \infty$. Here, $\int \mathd
\Gamma \cdot$ denotes a phase space integral and $A \left( \Gamma \right)$ is
a function of the phase space coordinate $\Gamma$. Choosing for $\mathsf{A}$
the identity $\mathsf{I}$, and for $\mathsf{Q}$ the solitonic solution
$\mathsf{P}_{0, 0}$ of (\ref{eq:NGCD}) with vanishing position and momentum
expectations, we obtain a resolution of the identity in terms of the solitons
$\mathsf{P}_{\Gamma} = \mathsf{T}_{s, u} \mathsf{P}_{0, 0} \mathsf{T}_{s,
u}^{^{\dag}}$, 
\begin{eqnarray}
  \mathsf{I} & = & \int \mathd \Gamma I \left( \Gamma \right) \bignone
  \mathsf{P}_{\Gamma} \, .  \label{eq:resolutionidentity}
\end{eqnarray}
This demonstrates that the pointer states of collisional decoherence form an
overcomplete basis.{\tmstrong{}}{\tmstrong{}}{\tmstrong{}}

\subsection{Dynamics in an external potential}

So far, we have characterized the solitonic solutions of the nonlinear
equation (\ref{eq:NGCD}) which applies in the absence of an external force. If
an additional potential is present the corresponding nonlinear equation
contains an additional term $V \left( x \right) / \left( i \hbar \right)$ on
the right hand side of (\ref{eq:NGCD}). The numerical treatment shows that the
solutions still converge to localized wave packets, which, however, change
their shape and velocity in the course of the evolution. We find that the
center of these wave packets moves on the corresponding classical phase space
trajectory provided the collision rate is sufficiently large. We first summarize our
numerical findings and then proceed with an analytic explanation.

Figure 5 shows the position and momentum expectation values of the numerical
solution of the nonlinear equation, in case of an anharmonic external
potential of the form $V \left( x \right) = ax^4 - bx^2, a, b > 0 \,$
(starting from an Gaussian initial state). The panel on the left hand side of
Fig.~5 was obtained in the limit of a vanishing collision rate $\gamma \,$
(i.e. $\kappa \rightarrow \infty$), which turns (\ref{eq:dimensionless}) into
the Schr\"odinger equation. The solution therefore disperses, and the solid
line shows a typical evolution of the phase space expectation values. The
dashed line, on the other hand, gives the classical trajectory of the phase
space point where the initial state is localized. The result for a large
collision rate $\gamma$ (or small $\kappa$) is shown on the right hand side of
Fig.~5. Here, the initial state turns rapidly into a soliton whose expectation
values move on the corresponding classical trajectory. This illustrates that
the temporal evolution turns from quantum to classical dynamics with increasing
collision rate $\gamma$ (i.e. decreasing $\kappa$). We made similar observations
with various other potentials. \ \ \ \ \ \

\begin{figure}[tb]

  \begin{center}
    \resizebox{12.5cm}{!}{\includegraphics{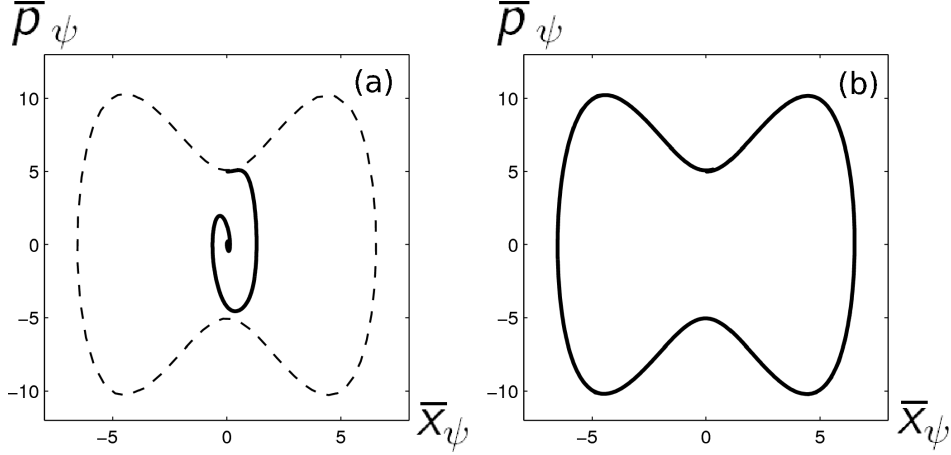}}
  \end{center}
  \caption{Time evolution of pointer states in an anharmonic fourth order
  potential (solid line). The dashed line shows the corresponding classical
  phase space trajectory. (a) The collision rate $\gamma$ vanishes leading to
  dispersive quantum dynamics. (b) The collision rate $\gamma$ is large, such
  that the dynamics of the pointer state is indistinguishable from the
  classical trajectory. \ \ \ \ \ }
\end{figure}

In order to explain the numerical observation, first consider a particle in a
linear potential $V \left( x \right) = \alpha x$. The corresponding nonlinear
equation reads as
\begin{eqnarray}
  \partial_t \psi_t \left( x \right) \text{} \text{} + \frac{\hbar^{}}{2 mi}
  \partial_x^2 \psi_t \left( x \right) & = &  \frac{1}{i \hbar} \alpha x
  \psi_t \left( x \right) + \gamma \psi_t \left( x \right) \Bigl[ | \psi_t |^2
  \ast \tilde{G} \left( x \right) \nonumber\\
  &  &  - \int^{\infty}_{- \infty} \mathd y| \psi_t |^2 \left( y
  \right)  \left( | \psi_t |^2 \ast \hat{G} \right) \left( y \right) \Bigr]
  \, .  \label{eq:linear}
\end{eqnarray}
As discussed in Sect.~4.1, the field-free version of this equation $(\alpha =
0)$ exhibits uniformly moving solitonic solutions of the form
\begin{eqnarray}
  \psi_t \left( x \right) & = & f \left( x - vt \right) \exp \left( i \left[
  \phi \left( x - vt \right) + \chi \left( t \right) \right] \right) \, . 
  \label{eq:movingsoliton}
\end{eqnarray}
This implies that (\ref{eq:linear}) has solitonic solutions of the form
\begin{eqnarray}
  \psi_t \left( x \right) & = & f \left( x - x_t \right) \exp \left( ig \left(
  x - x_t, t \right) \right) \,,  \label{eq:accelerated}\\
  g \left( x, t \right) & = & \phi \left( x \right) + \chi' \left( t \right) -
  \frac{\alpha}{\hbar} tx \,,  \label{eq:definitiong}
\end{eqnarray}
with $x_t = vt - \alpha t^2 / 2 m \,$ and $\chi' \left( t \right) = \chi
\left( t \right) - 2 \alpha \int_0^t \mathd \tau \, x_{\tau} / \hbar
\bignone$. In order to verify this statement, we evaluate the left hand side
of (\ref{eq:linear}) with the ansatz given by (\ref{eq:accelerated}). This
yields
\begin{eqnarray}
 \fl \partial_t \psi_t \left( x \right)_{} + \frac{\hbar}{2 mi} \partial_x^2
  \psi_t \left( x \right) & = & e^{ig} \left( \frac{\alpha}{i \hbar} xf + if
  \partial_t \chi' \left( t \right) + \frac{2 \alpha}{\hbar} ix_t f - v \left(
  \partial_x f + if \partial_x \phi \right) \right. \nonumber\\
  &  & \left. + \frac{\hbar}{m} \left( \partial_x f \partial_x \phi -
  \frac{i}{2} \partial_x^2 f + \frac{i}{2} f \left( \partial_x \phi \right)^2
  + \frac{1}{2} f \partial_x^2 \phi \right) \right) \,, 
  \label{eq:derivatives}
\end{eqnarray}
with $g \equiv g \left( x - x_t, t \right)$, $f_{} \equiv f \left( x - x_t
\right)$ and $\phi_{} \equiv \phi \left( x - x_t \right)$. The expression can
be further simplified by noting that the free soliton (\ref{eq:movingsoliton})
is a solution of the field-free version of (\ref{eq:linear}), implying that
\begin{eqnarray}
  \gamma f \Lambda \left[ f^2 \right] \left( x \right) & = & if \partial_t
  \chi \left( t \right) - v \left( \partial_x f + if \partial_x \phi \right)
  \nonumber\\
  &  & + \frac{\hbar}{m} \left( \partial_x f \partial_x \phi - \frac{i}{2}
  \partial_x^2 f + \frac{i}{2} f \left( \partial_x \phi \right)^2 +
  \frac{1}{2} f \partial_x^2 \phi \right) \,,  \label{eq:freefieldcase}
\end{eqnarray}
with $\Lambda \left[ f^2 \right] \left( x \right)$ defined in
(\ref{eq:lambda}). Using (\ref{eq:derivatives}), (\ref{eq:freefieldcase}) and
the above definition of $\chi' \left( t \right)$, one finds that
\begin{eqnarray}
  \partial_t \psi_t \left( x \right) + \frac{\hbar}{2 mi} \partial_x^2 \psi_t
  \left( x \right) & = & \frac{1}{i \hbar} \alpha x \psi_t \left( x \right) +
  \gamma \psi_t \left( x \right) \Lambda \left[ \left| \psi_t \right|^2
  \right] \left( x \right) \,,
\end{eqnarray}
which confirms that $\psi_t \left( x \right)$ evolves according to
(\ref{eq:linear}).

We conclude that in a linear potential the pointer states have the same shape
as in the field-free case and they are uniformly accelerated like a classical
particle. For general potentials, this implies that the pointer states follow
the corresponding classical motion, provided the spatial width of the solitons
is sufficiently small, such that the linearization of the potential is
justified over their spatial extension. Since the size of the pointer states
decreases with the collision rate (see Section 4.3), the pointer states must
exhibit classical dynamics in the limit of large collision rates.

\section{Orthogonal Unraveling}

Thus far, we have calculated `candidate' pointer states as the solitonic
solutions of (\ref{eq:NGCD}), and we have studied their properties and
dynamics. Next, it will be shown that these `candidates' are genuine pointer states
$\mathsf{P}_{\alpha}$ in the sense of (1). Moreover, we find that the
statistical weights $\tmop{Prob} \left( \alpha | \rho_0 \right)$ of the
pointer states are given by the overlap of the initial state $\rho_0$ with the
initial pointer state $\mathsf{P}_{\alpha} \left( 0 \right)$, i.e.
\begin{eqnarray}
  \tmop{Prob} \left( \alpha | \rho_0 \right) & = & \tmop{Tr} \left[ \rho_0
  \mathsf{P}_{\alpha} \left( 0 \right) \right] \, .  \label{eq:initaloverlap}
\end{eqnarray}
In order to verify the above conjectures, let us now make use of the formalism of
quantum trajectories {\cite{Diosi1986a,Carmichael1993a,Breuer2007b}} to solve
the master equation (\ref{eq:CD}). More precisely, a specific quantum
trajectory method, the {\tmem{orthogonal unraveling}}, is distinguished by the
physics of pointer states because the deterministic part of the associated
stochastic differential equation coincides with the nonlinear equation
(\ref{eq:NG}).

We start with a general description of the quantum trajectory approach and
the orthogonal unraveling in Section 5.1. The latter will be applied to
collisional decoherence in Section 5.2. This will allow us to evaluate the
statistical weights of the pointer states in Section 5.3.

\subsection{Quantum trajectories and the orthogonal unraveling}

\subsubsection{Quantum trajectories }

In the quantum trajectory approach, the wave function corresponding to a pure
initial state $| \psi \left( 0 \right) \rangle$ is propagated stochastically
to generate pure state trajectories $\left\{ | \psi_i \left( t \right) \rangle
\right\}$, whose ensemble average recovers the solution of the master equation
(\ref{eq:lindblad}), i.e.
\begin{eqnarray}
  \exp \left( \mathcal{L}t \right) | \psi \left( 0 \right) \rangle \langle
  \psi \left( 0 \right) | & = & \mathbbm{E} \left( | \psi_i \left( t \right)
  \rangle \langle \psi_i \left( t \right) | \bignone \right) \, . 
  \label{eq:average}
\end{eqnarray}
Such a stochastic process is called an unraveling of the master equation. A
common unraveling is provided by the quantum Monte Carlo method
{\cite{Molmer1993a,Molmer1996a}} which is based on a piecewise deterministic
process. Here, one realization of a trajectory consists of smooth
deterministic pieces generated by an effective (non-Hermitian) Hamiltonian, in
our case
\begin{eqnarray}
  \mathsf{H}_{\tmop{eff}} & = & \mathsf{H}_{} - \frac{i \hbar}{2} \int_{-
  \infty}^{\infty} \mathd q \bignone \mathsf{L}_q^{\dag} \mathsf{L}_q \,, 
  \label{eq:cont1}
\end{eqnarray}
which are interrupted by random jumps. The jumps occur with the rate
\begin{eqnarray}
  r_q & = & \langle \mathsf{L}_q^{\dag} \mathsf{L}_q \rangle \,,
  \label{eq:rate}  \label{eq:rate1}
\end{eqnarray}
an expectation value with respect to $| \psi_i \left( t \right) \rangle$, and they
are effected by the operators
\begin{eqnarray}
  \mathsf{J}_q & = & \mathsf{L}_q / \sqrt{r_q} \, .  \label{eq:jumpmc}
\end{eqnarray}
The quantum Monte Carlo method is not the only stochastic process satisfying
(\ref{eq:average}). In fact, there exists an infinite set of these stochastic
processes, because the convex decomposition of the density matrix on the left hand
side of (\ref{eq:average}) is not unique. For instance, other unravelings can
be obtained from the quantum Monte Carlo method, by noting that the generator
$\mathcal{L}$ does not uniquely fix the Lindblad operators $\mathsf{L}_q \,$
and the Hamiltonian $\mathsf{H}$ {\cite{Molmer1996a,Breuer2007b}}. This is due
to the fact that the master equation (\ref{eq:lindblad}) is invariant under
certain transformations of the Lindblad operators, such as the addition of a
complex multiple $z_q$ of the identity
$\mathsf{\mathsf{\mathsf{\mathsf{I}}}}$,
\begin{eqnarray}
  \mathsf{L}_q & \rightarrow & \mathsf{L}_q' = \mathsf{L}_q + z_q \mathsf{I}
  \, .  \label{eq:trafo1}
\end{eqnarray}
In the latter case, also the Hamiltonian must be transformed as
\begin{eqnarray}
  \mathsf{H} & \rightarrow & \mathsf{H}' = \mathsf{H} + \frac{1}{2 i} \int_{-
  \infty}^{\infty} \mathd q \bignone \left( z_q^{\ast} \mathsf{L}_q - z_q
  \mathsf{L}_q^{\dag} \right) \,,  \label{eq:trafo2}
\end{eqnarray}
in order to assure the invariance of the master equation.

\subsubsection{The orthogonal unraveling}

To obtain a different, but again piecewise deterministic unraveling, we now
make the choice $z_q = - \langle \mathsf{L}_q \rangle$. It then follows that
the deterministic pieces of a sample path are given by the solution of the
nonlinear equation (\ref{eq:nonlinear}) (or equivalently the corresponding
projector equation (\ref{eq:NG})). The jumps occur with the rate
\begin{eqnarray}
  r_q & = & \langle \mathsf{L}_q^{\dag} \mathsf{L}_q \rangle - \langle
  \mathsf{L}_q^{\dag} \rangle \langle \mathsf{L}_q \rangle \,, 
  \label{eq:rate2}
\end{eqnarray}
and are caused by the nonlinear operators
\begin{eqnarray}
  \mathsf{J}_q & = & \left( \mathsf{L}_q - \langle \mathsf{L}_q \rangle
  \right) / \sqrt{r_q} \, .  \label{eq:jump}
\end{eqnarray}
As a distinctive feature, the states $| \psi_q \left( t \right) \rangle =
\mathsf{J}_q | \psi_{} \left( t \right) \rangle$ into which the system may
jump are orthogonal to the original state $| \psi \left( t \right)
\rangle$, thus justifying its naming. (The
states $| \psi_q \left( t \right) \rangle$ are not necessarily mutually
orthogonal, though.) To our knowledge, this unraveling was first noted by Rigo
and Gisin {\cite{Rigo1996a}}, although, it has not been studied numerically so
far.

A related unraveling, which is also referred to as the `orthogonal
unraveling', was introduced by Di\'osi {\cite{Diosi1985a,Diosi1986a}}. Here, the
deterministic pieces of the evolution are as well generated by
(\ref{eq:nonlinear}). However, the states $| \psi_q \left( t \right) \rangle$
into which the system may jump are obtained differently, as the eigenvectors
of the Hermitian operator
\begin{eqnarray}
  \left( 1 - \mathsf{P}_{\psi_{} \left( t \right)} \right) \mathcal{L} \left(
  \mathsf{P}_{\psi_{} \left( t \right)}) \left( 1 - \mathsf{P}_{\psi_{} \left(
  t \right)} \right) \,, \right.  \label{eq:rateoperator}
\end{eqnarray}
where $\mathsf{P}_{\psi_{} \left( t \right)} \equiv | \psi_{} \left( t \right)
\rangle \langle \psi_{} \left( t \right) | $. As a consequence, these states
are also {\tmem{mutually}} orthogonal (in finite dimensional systems).
Since the orthogonal unraveling of {\cite{Diosi1985a,Diosi1986a}} requires
the diagonalization of the operator (\ref{eq:rateoperator}), it is much more
involved than the one defined by (\ref{eq:nonlinear}), (\ref{eq:rate2}) and
(\ref{eq:jump}), which is why we will use the latter in the following.

\subsubsection{Pointer states and the orthogonal unraveling }

As mentioned above, all unravelings are equivalent in the same sense as the
different convex decompositions of the density matrix $e^{\mathcal{L}t}
\rho_0$. Note, however, that a preferred set of pure states -- the pointer
basis -- may be singled out through the environmental coupling. In that case,
those unravelings are distinguished which generate for any initial state an ensemble of these state independent projectors. An unraveling will do this job if (a)
its deterministic part exhibits stable fixed points or solitons
$\mathsf{P}_{\alpha} = | \pi_{\alpha} \rangle \langle \pi_{\alpha} |$, which
(b) are characterized by a vanishing jump rate, $r \left( \mathsf{P}_{\alpha}
\right) = 0 \,$. In that case, the sample paths of the process will end up in
one of the states $\mathsf{P}_{\alpha}$, by all means. Hence, the ensemble
mean is of the form (\ref{eq:decoherence}), such that the fixed points or
solitons $\mathsf{P}_{\alpha}$ can be identified with the pointer states of
the system. \ \

For the case of collisional decoherence the orthogonal unraveling fulfills
the aforesaid conditions. (a) Its deterministic part is given by
(\ref{eq:NGCD}) which exhibits the solitonic solutions $\pi \left( x \right)$
shown in Fig.~2. We shall see explicitly in Section 5.2.1 that these states
are attractive fixed points. (b) We will show in Section 5.2.2 that the jump
rate (\ref{eq:rate2}) vanishes for the solitons $\pi \left( x \right)$, which
finally demonstrates that our `candidate' pointer states $\pi \left( x
\right)$ are genuine pointer states $\mathsf{P}_{\alpha}$ in the sense of (1).
Moreover, this shows that the orthogonal unraveling is a very efficient
numerical scheme for the long time solution of the master equation
(\ref{eq:CD}), since the state is no longer affected by the stochastic part,
once it has turned into the soliton, and the trajectory is therefore more easy
to integrate.

We note that the orthogonal unraveling is not the only stochastic process
which generates the ensemble of pointer states. In fact, there is a
{\tmem{diffusive}} unraveling {\cite{Gisin1992a}}, which also involves
(\ref{eq:NG}) as its deterministic part. It was applied in {\cite{Diosi2000a}}
to investigate pointer states in a linear model.

\subsection{Unraveling collisional decoherence }

We now apply the described orthogonal unraveling to collisional decoherence
(\ref{eq:CD}), first evaluating the deterministic part (\ref{eq:nonlinear}) of
the stochastic process in Sect.~5.2.1 and then the stochastic one
(\ref{eq:rate2}) and (\ref{eq:jump}) in Sect.~5.2.2.

\subsubsection{Deterministic evolution}

Applying (\ref{eq:nonlinear}) to the case of collisional decoherence yields
the soliton equation (\ref{eq:NGCD}) discussed in Sect.~3.2. We will now
further simplify this equation, by considering initial states
\begin{eqnarray}
  \Psi_0 \left( x \right) & = & \sum_{i = 1}^N c_i \left( 0 \right) \phi_i
  \left( x, 0 \right) \,,  \label{eq:initialstate}
\end{eqnarray}
which are superpositions of non-overlapping wave functions $\phi_i \left( x, 0
\right)$,
\begin{eqnarray}
  \phi_i \left( x, 0 \right) \phi_{j \neq i}^{\ast} \left( x, 0 \right) & = &
  0 \, .  \label{eq:A1}
\end{eqnarray}
The latter are assumed to be localized in the sense that
\begin{eqnarray}
  \sigma^2_{\phi_i} & < & \frac{2 \pi \hbar^2}{\sigma_G^2} \,,  \label{eq:A2a}
\end{eqnarray}
where $\sigma^2_{\phi_i}$ and $\sigma_G^2$ denote the variances of the
distributions $\left| \phi_i \left( x, 0 \right) \right|^2$ and $G \left( q
\right)$, respectively. Under this assumption, which will be justified at the
end of this section, one can extract a system of evolution equations for the
time evolution of the coefficients in (\ref{eq:initialstate}),
\begin{equation}
  \frac{\mathd}{\mathd t} c_i \left( t \right)  =  - \left( \sum_{j = 1}^N
  F_{ij} \left| c_j \left( t \right) \right|^2 - \sum_{j, k = 1}^N F_{jk}
  \left| c_j \left( t \right) \right|^2 \left| c_k \left( t \right) \right|^2
  \bignone \bignone \right) c_i \left( t \right) \, .  \label{eq:coefficients}
\end{equation}
Here, the matrix $F_{ij} \equiv F \left( x_i - x_j \right)$ is obtained from
the localization rate (\ref{eq:localization}), where the $x_i \equiv \langle
\mathsf{x} \rangle_{\phi_i}$ denote the mean positions of the constituent wave
functions $\phi_i \left( x, t \right)$. The latter evolve according to
\begin{eqnarray}
  \partial_t \phi_i \left( x, t \right) & = & - \frac{\hbar^{}}{2 \, m \, i}
  \partial_x^2 \phi_i \left( x, t \right) + \phi_i \left( x, t \right) \Lambda
  \left[ \left| \phi_i \right|^2 \right] \left( x, t \right) \nonumber\\
  &  & + \phi_{^{} i} (x, t) \sum^N_{j = 1, j \neq i} \bignone \left| c_j
  \left( t \right) \right|^2  \tilde{\gamma}_{ij} \left( x, t \right) \,, 
  \label{eq:basis}
\end{eqnarray}
where $\Lambda$ is defined in (\ref{eq:lambda}) and $\tilde{\gamma}_{ij}$ is a
rate of the order of $\gamma$,
\begin{eqnarray}
  \tilde{\gamma}_{ij} \left( x, t \right) & \equiv & \left| \phi_i \right|^2
  \ast F \left( x, t \right) - \left| \phi_j \right|^2 \ast F \left( x, t
  \right) + F_{ij} \, .  \label{eq:basisrate}
\end{eqnarray}
Let us now verify that $\Psi_t \left( x \right) = \sum_{i = 1}^N c_i \left( t
\right) \phi_i \left( x, t \right)$, with $c_i \left( t \right)$ and $\phi_i
\left( x, t \right)$ solutions of (\ref{eq:coefficients}) and
(\ref{eq:basis}), evolves according to (\ref{eq:NGCD}). First, we note that
the assumption (\ref{eq:A2a}) leads to the approximation
\begin{eqnarray}
  \int_{- \infty}^{\infty} \mathd x \left| \phi_i \left( x \right) \right|^2
  e^{iqx / \hbar} \bignone & \simeq & e^{iqx_i / \hbar} \,,  \label{eq:A2b}
\end{eqnarray}
for all $q$ contributing appreciably to integrals weighted with the momentum
transfer distribution $G \left( q \right)$. This, in turn, implies
\begin{eqnarray}
  F_{jk} & = & F \left( x_j - x_k \right)  \label{eq:Fij}\\
  & \simeq & \int_{- \infty}^{\infty} \mathd x \left| \phi_j \left( x \right)
  \right|^2 \left( \left| \phi_k \right|^2 \ast F \right) \left( x \right) \,
  . \nonumber
\end{eqnarray}
Hence, we find that $\int_{- \infty}^{\infty} \mathd x \left| \phi_i \left( x
\right) \right|^2 \left( \left| \phi_i \right|^2 \ast F \right) \left( x
\right) \simeq 0$, and one thus obtains
\begin{eqnarray}
  \Lambda \left[ \left| \phi_i \right|^2 \right] \left( x, t \right) & = & -
  \left( \left| \phi_i \right|^2 \ast F \right) \left( x, t \right) \, . 
  \label{eq:lambdaA2}
\end{eqnarray}
Now, consider the time derivative of $\Psi_t \left( x \right)$, which gives
\
\begin{eqnarray}
  \partial_t \Psi_{} & = & - \frac{\hbar}{2 \, m \, i} \sum_{i = 1}^N c_i
  \bignone \partial_x^2 \phi_i - \left( \sum_{i = 1}^N c_i \phi_i \right) \,
  \nonumber\\
  &  & \times \left( \sum_{j = 1}^N \bignone \bignone \left| c_j \right|^2
  \left( \left| \phi_j \right|^2 \ast F \right) - \sum_{j, k = 1}^N \left| c_j
  \right|^2 \left| c_k \right|^2 F_{jk} \bignone \right) \,, 
  \label{eq:beweis1}
\end{eqnarray}
where we dropped the arguments for brevity. In (\ref{eq:beweis1}), we used
(\ref{eq:coefficients})-(\ref{eq:basisrate}), (\ref{eq:lambdaA2}), the fact
that $F_{ii} = 0$, and the normalization condition $\sum_{i = 1}^N \left| c_i
\right|^2 = 1 \bignone$. Now, we replace $F_{jk}$ in (\ref{eq:beweis1}) by the
right hand side of (\ref{eq:Fij}), and use (\ref{eq:A1}), which yields
\begin{eqnarray}
  \partial_t \Psi_t \left( x \right)_{} & = & - \frac{\hbar}{2 mi}
  \partial_{x^{}}^2 \Psi_t \left( x \right) - \Psi_t \left( x \right)  \Bigl(
  | \Psi_t |^2 \ast F \left( x \right)  \nonumber\\
  &  &  - \int_{- \infty}^{\infty} \mathd x| \Psi_t |^2 \left( y
  \right) \left( | \Psi_t |^2 \ast F \right) \left( y \right) \Bigr)\, . 
  \label{eq:nonlinearequation1}
\end{eqnarray}
Finally, by using (\ref{eq:localization}), we obtain
\begin{eqnarray}
 \partial_t \Psi_t \left( x \right) & = & - \frac{\hbar^{}}{2 mi}
  \partial_x^2 \Psi_t + \gamma \Psi_t \left( x \right)  \label{eq:nonlinearequation2}\\
  &  &  \times \left( | \Psi_t |^2 \ast \hat{G}
  \left( x \right) - \int^{\infty}_{- \infty} \mathd y| \Psi_t |^2 \left( y
  \right)  \left( | \Psi_t |^2 \ast \hat{G} \right) \left( y \right) \right)
  \,, \nonumber
\end{eqnarray}
which confirms that $\Psi_t \left( x \right)$ evolves according to
(\ref{eq:NGCD}).

\begin{figure}[tb]

  \begin{center}
    \resizebox{13cm}{!}{\includegraphics{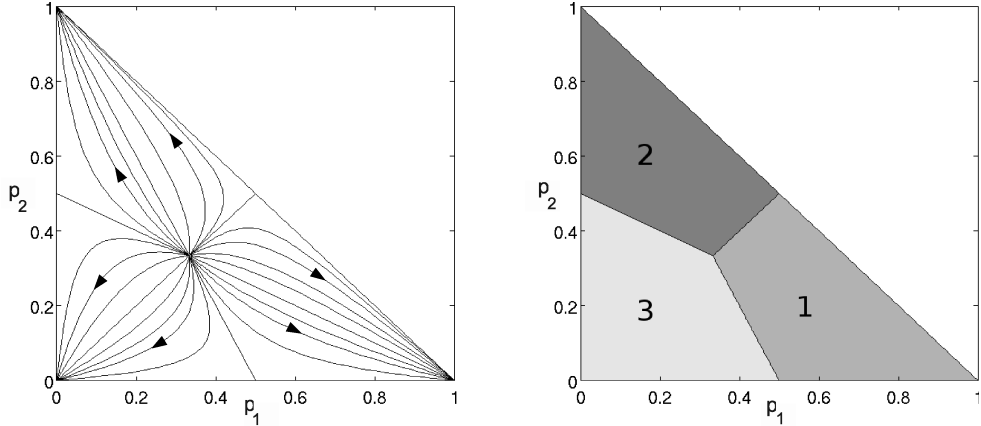}}
  \end{center}
  \caption{Numerical solution of (\ref{eq:coefficients}) for $N = 3$. The
  $x$-axis gives $p_1 = \left| c_1 \right|^2$, the $y$-axis $p_2 = \left|
  c_2 \right|^2$, and $\left| c_3 \right|^2$ is fixed by normalization. Left:
  trajectories indicating the flow into the stable fixed points $\left| c_i
  \right| = \delta_{i, n}, \left( n = 1, 2, 3 \right)$. Right: regions of
  attraction of the stable fixed points; the area denoted with $n$ is the
  region of attraction of the fixed point $\left| c_i \right| = \delta_{i, n}
  \, .$ \ \ }
\end{figure}

Let us now discuss the solution of Eq.~(\ref{eq:coefficients}) for the time
evolution of the coefficients. We first consider situations where the wave
packets $\phi_i \left( x \right)$ are sufficiently far apart such that the
localization rate is saturated, i.e. $F_{ij} = \gamma \left( 1 - \delta_{ij}
\right) .$ Under this assumption, (\ref{eq:coefficients}) reduces to the
equation
\begin{eqnarray}
  \frac{\mathd}{\mathd t} c_i \left( t \right) & = & - \gamma \left( \sum_{j =
  1}^N \left| c_j \left( t \right) \right|^4 - |c_i \left( t \right) |^2
  \bignone \bignone \right) c_i \left( t \right) \,, \bignone 
  \label{eq:coefficientssaturated}
\end{eqnarray}
which was already studied in {\cite{Diosi1988a}} in the context of a discrete
model for quantum measurement. It is shown there that all stable fixed points
of (\ref{eq:coefficientssaturated}) have the form $\left| c_i \right| =
\delta_{i, n} \,$, and that the particular fixed point $\left| c_i \right| =
\delta_{i, m} \,$, with
\begin{eqnarray}
  m & = & \text{arg$\max_i \left( \left| c_i \left( 0 \right) \right|^2
  \right)$\,,}  \label{eq:argmax}
\end{eqnarray}
is approached monotonically, i.e. the component with the largest initial
weight wins. This behavior is visualized in Fig.~6 which was obtained by
solving (\ref{eq:coefficientssaturated}) numerically for the case $N = 3$.
Here, the $x$- and the $y$-axis indicate the weights $p_1 = \left| c_1
\right|^2$ and $p_2 = \left| c_2 \right|^2$, respectively. The plot on the
left hand side shows various trajectories, illustrating in particular the
fixed points. The plot on the right displays the regions of attraction of the
stable fixed points $\left| c_i \right| = \delta_{i, n}$, in agreement with
the criterion (\ref{eq:argmax}). For instance, area $1$ highlights the region
of attraction of the fixed point $\left| c_i \right| = \delta_{i, 1}$. \ \ \ \

Figure 7, on the other hand, depicts a scenario where the wave packets
$\phi_i \left( x \right)$ are close together such that the localization rate
is unsaturated, i.e. $F_{ij} \leqslant \gamma \left( 1 - \delta_{ij} \right)$.
Here, we choose  $N = 3$ wave packets with non-equidistant position expectations, $\left(
x_1, x_2, x_3 \right) \sigma_G / \hbar = \left( 1.4, 1.3, 0.8 \right)$. Similarly to the saturated case, we observe that  all stable fixed points of
(\ref{eq:coefficients}) have the form $\left| c_i \right| = \delta_{i, n} \,$.
However, the regions of attraction are deformed such that criterion
(\ref{eq:argmax}) is no longer valid, and the fixed points are not necessarily
approached monotonically. \ \

To see that $\left| c_i \right| = \delta_{i, n} \,$ are stable fixed points
of (\ref{eq:coefficients}), assume that the coefficients are close by, i.e.
$\left| c_n \left( t \right) \right| = 1 - \varepsilon \left( t \right)$ with
$\varepsilon \left( 0 \right) \ll 1$. It follows from (\ref{eq:coefficients})
that
\begin{eqnarray}
  \dot{\varepsilon} \left( t \right) & = & - \sum_j F_{nj} \, \left| c_j
  \left( t \right) \right|^2 + O \left( \varepsilon^2 \right) \: \bignone < \:
  0 \,,  \label{eq:timederivativeepsilon}
\end{eqnarray}
and hence, $\left| c_i \left( t \rightarrow \infty \right) \right| =
\delta_{i, n}$.

The knowledge of the fixed points of the coefficients allows one to discuss
the asymptotic evolution of the initial state shown in
Eq.~(\ref{eq:initialstate}). Since the coefficients $c_j$ with $j \neq m \,$
tend to zero asymptotically, it follows that
\begin{eqnarray}
  | \psi \left( t \rightarrow \infty \right) \rangle & = & | \phi_m \left( t
  \rightarrow \infty \right) \rangle \,,  \label{eq:asymptoticstate}
\end{eqnarray}
for a specific $m$ (which is given by (\ref{eq:argmax}) in the saturated
case). The asymptotic behavior of the wave packets $| \phi_m \rangle$ can, in
turn, be predicted from Eq.~(\ref{eq:basis}). Since the $c_{j \neq m}$ vanish
for large times, the coupling term given by the last summand in
{\tmstrong{}}(\ref{eq:basis}) vanishes as well, implying that the time
evolution (\ref{eq:basis}) of $| \phi_m \rangle$ is asymptotically equal to
the soliton equation (\ref{eq:NGCD}). Therefore, in the absence of stochastic
jumps, the state $| \Psi_0 \rangle$ evolves into that solitonic solution $\pi_m
\left( x \right)$ of (\ref{eq:NGCD}) which is associated to the initial wave
packet $| \phi_m \left( 0 \right) \rangle$.

It should be mentioned that Eqs.~(\ref{eq:coefficients}) and
{\tmstrong{}}(\ref{eq:basis}) for the coefficients $c_i$ and the constituent
wave packets $\phi_i$ are not completely decoupled, since
(\ref{eq:coefficients}) depends on the matrix $F_{ij} \equiv F \left( x_i -
x_j \right)$ which contains the position expectations $x_i$ of the wave
packets $\phi_i$. However, the position expectation follows the classical
trajectory for sufficiently large $\kappa$'s, such that
(\ref{eq:coefficients}) can be solved without knowing the solution of
{\tmstrong{}}(\ref{eq:basis}).

\begin{figure}[tb]

  \begin{center}
    \resizebox{13.5cm}{!}{\includegraphics{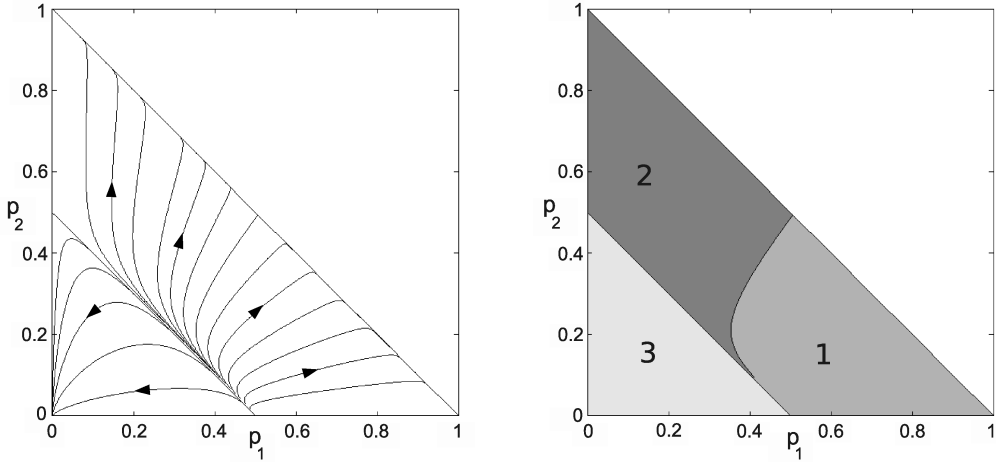}}
  \end{center}
  \caption{Similar to Fig.~6, but the wave packets $\phi_i \left( x \right)$
  are positioned closer such that the localization rate is unsaturated, i.e.
  $F_{ij} \leqslant \gamma \left( 1 - \delta_{ij} \right)$. The stable fixed
  points are still of the form $\left| c_i \right| = \delta_{i, n}$, but they
  may be approached non-monotonically and the regions of attraction, shown on
  the right, are deformed compared to the saturated case.}
\end{figure}

Let us now discuss the validity of the assumption of small position variance
(\ref{eq:A2a}), and the ensuing approximation (\ref{eq:A2b}). It can be
justified by our observation in Section~4.3, that the dimensionless pointer
width $\sigma_{\pi} \sigma_G / \hbar$ is a function of the parameter $\kappa
\equiv \sigma_G^2 / m \hbar \gamma$ only,
\begin{eqnarray}
  \sigma_{\pi} \frac{\sigma_G}{\hbar} & = & \frac{\kappa}{4 a_{\tmop{loc}}} +
  a_{\tmop{loc}} \,, \hspace{2em} \tmop{with} \,\, a_{\tmop{loc}} =
  0.4 \, .  \label{eq:widthofsoliton}
\end{eqnarray}
Thus, for all $\kappa \ll 4 a^2_{\tmop{loc}} \simeq 1$ we find that the
position variance $\sigma_{\pi}^2$ of a pointer state is one order of
magnitude smaller than the reciprocal width of the momentum transfer
distribution $G \left( q \right)$,
\begin{eqnarray}
  \text{$\sigma_{\pi}^2$} \simeq a_{\tmop{loc}}^2 \frac{\hbar^2}{\sigma_G^2}
  \simeq 0.2 \frac{\hbar^2}{\sigma_G^2} < 2 \pi \frac{\hbar^2}{\sigma_G^2} \,
  . &  &  \label{eq:justificationvarianceassumption}
\end{eqnarray}
The above relation for the width of the pointer state is sufficient to justify
the approximation $\int \mathd x \bignone e^{iqx / \hbar} \left| \pi \left( x
\right) \right|^2 \simeq e^{iq \langle \mathsf{x} \rangle_{\pi} / \hbar}$, as
we checked numerically, by using the solitonic solution of (6). The relative
error is less than $2\%$ for $q \in \left[ - 2 \sigma_G, 2 \sigma_G \right]$
and $\kappa \leqslant 10^{- 3}$.

\subsubsection{Stochastic part}

Upon inserting the Lindblad operator $\mathsf{L}_q = \sqrt{\gamma G \left( q
\right)} e^{iq \mathsf{x}}$ into (\ref{eq:jump}) we find that the jump
operator takes the form
\begin{eqnarray}
  \mathsf{J}_q & = & \mathcal{N}_q \, \left( e^{iq \mathsf{x} / \hbar} -
  \langle e^{iq \mathsf{x} / \hbar} \rangle \right) \,, 
  \label{eq:jumpoperator}
\end{eqnarray}
with normalization $\mathcal{N}_q = \left( 1 - | \langle e^{iq \mathsf{x} /
\hbar} \rangle |^2 \right)^{- 1 / 2}$. We again consider states of the form
(\ref{eq:initialstate}) which are superpositions of non-overlapping
(\ref{eq:A1}) and localized (\ref{eq:A2a}) wave packets $\phi_i \left( x
\right)$. Under this assumption, one can evaluate the expectation value in
(\ref{eq:jumpoperator})
\begin{eqnarray}
  \langle e^{iq \mathsf{x} / \hbar} \rangle_{\Psi} & = & \sum_{i = 1}^N \left|
  c_j \right|^2 e^{iqx_j / \hbar} \,,  \label{eq:expectation}
\end{eqnarray}
such that the state $\Psi_q \left( x \right) \equiv \mathsf{J}_q \Psi \left( x
\right)$ into which the system may jump takes the form
\begin{eqnarray}
  \Psi_q \left( x \right) & = & \mathcal{N}_q \left( e^{iqx_{} / \hbar} -
  \sum_{i = 1}^N \left| c_j \right|^2 e^{iqx_j / \hbar} \right) \sum_{i = 1}^N
  c_i \, \phi_i \left( x \right) \, .  \label{eq:stateafterjump}
\end{eqnarray}

Later we will choose the initial wave packets $\phi_i \left( x \right)$ to be
solitons $\pi_i \left( x \right)$. Let us therefore assume that the $\phi_i$'s
form a basis, such that $\Psi_q \left( x \right)$ can be represented as
$\Psi_q \left( x \right) = \sum_i c_i \left( q \right) \phi_i \left( x \right)
\, \bignone$. Then the transformed coefficients $c_k \left( q \right)$ can be
evaluated by the overlap $c_k \left( q \right) = \langle \phi_k | \mathsf{J}_q
| \Psi \rangle$. Using (\ref{eq:A1}) and (\ref{eq:A2b}) this leads to the
following expression for the redistribution of the coefficients due to an
orthogonal jump
\begin{eqnarray}
  c_k \left( q \right) & = & \mathcal{N}_q \left( e^{iqx_k / \hbar} - \sum_{i
  = 1}^N \left| c_i \right|^2 e^{iqx_i / \hbar} \right) c_k \,, 
  \label{eq:process2}
\end{eqnarray}
Similarly, one can evaluate the rate (\ref{eq:rate2}) associated to the jump
operator of collisional decoherence,
\begin{eqnarray}
  r_q & = & \gamma \, G \left( q \right) \left( 1 - | \langle e^{iq \mathsf{x}
  / \hbar} \rangle |^2 \right) \, .  \label{eq:exactrate}
\end{eqnarray}
The above approximation (\ref{eq:expectation}) further simplifies this
expression,
\begin{eqnarray}
  r_q & = & \gamma \, G \left( q \right) \left( 1 - \sum_{j, k = 1}^N \left|
  c_j \right|^2 \left| c_k \right|^2 e^{iq \left( x_j - x_k \right) / \hbar}
  \bignone \right) \, .  \label{eq:process3}
\end{eqnarray}
One observes that this rate vanishes for the stable fixed points $\left| c_i
\right| = \delta_{i, n}$, indicating that the quantum trajectories of the
orthogonal unraveling evolve into the pointer states, the solitonic solutions
of (\ref{eq:NGCD}). Moreover, we note that the original stochastic process
(\ref{eq:nonlinear}), (\ref{eq:rate2}), (\ref{eq:jump}) -- which is defined in
the infinite dimensional Hilbert space of the system -- has been reduced to a
stochastic process in $\mathbbm{C}^N$, demonstrating the efficiency of the
orthogonal unraveling. However, due to the finite pointer width
(\ref{eq:sizedimensionless}) the {\tmem{exact}} expression for the jump rate
(\ref{eq:exactrate}) does not vanish identically, although it is very small
compared to $\gamma$. For instance, the numerically obtained soliton displays
a strongly suppressed total jump rate $r_{\tmop{tot}} = \int \mathd q \bignone
\, r_q$ of $r_{\tmop{tot}} / \gamma = 7 \times 10^{- 3}$ for $\kappa = 10^{-
3}$, while the superposition state decays with the rate $r_{\tmop{tot}} \cong
\gamma$. This implies that the solitons are not perfect pure state solutions
of the master equation (\ref{eq:CD}), though the loss of purity is small.

\subsection{The statistical weights of the pointer states}

The previous section showed that the orthogonal unraveling of an initial
superposition state subject to collisional decoherence can be reduced to a
stochastic process with respect to the corresponding coefficients. In
particular, this applies to the case where the initial state is a
superposition of pointer states,
\begin{eqnarray}
  | \Psi_0 \rangle & = & \sum_{i = 1}^N c_i | \pi_i \left( 0 \right) \rangle
  \,. \bignone  \label{eq:superpositionof*PS}
\end{eqnarray}
Thus we can now verify, by using the discrete process defined by
Eqs.~(\ref{eq:coefficients}), (\ref{eq:process2}) and (\ref{eq:process3}),
that after decoherence the statistical weights of the pointer states are given
by the overlap of the initial state with the initial pointer states. More
specifically, this demonstrates that the initial state $\Psi_0 \left( x
\right)$ evolves into the mixture
\begin{eqnarray}
  \rho \left( x, x' \right) & = & \sum_{i = 1}^N p_i \pi_i \left( x \right)
  \pi_i^{\ast} \left( x' \right) \,,  \label{eq:mixturePS}
\end{eqnarray}
where the statistical weights are given by the overlap
\begin{eqnarray}
  p_i & = & \left| \langle \Psi_0 | \pi_i \left( x, 0 \right) \rangle
  \right|^2 \, .  \label{eq:distribution}
\end{eqnarray}
We first present an analytic proof of the above for $N = 2 \,$. The general
case, $N > 2 \,$, is then treated numerically in the following section.

\begin{figure}[tb]

  \begin{center}
    \resizebox{9cm}{!}{\includegraphics{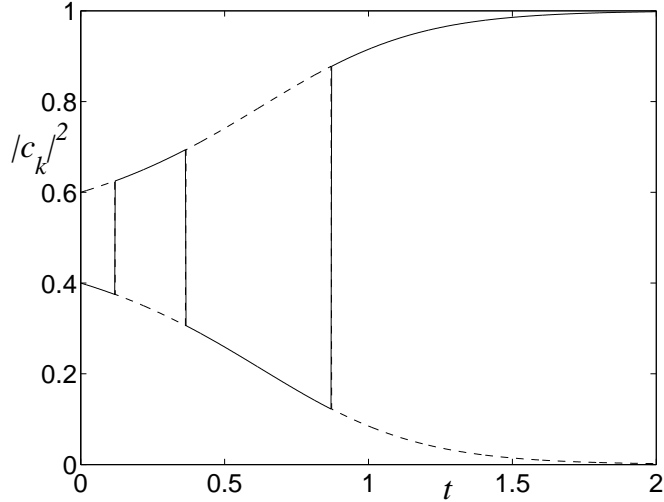}}
  \end{center}
  \caption{Quantum trajectory generated by (\ref{eq:coefficients}),
  (\ref{eq:averagejump}) and (\ref{eq:totalrate}) with $N = 2 \,$. The solid
  line depicts the evolution of $\left| c_1 \left( t \right) \right|^2 \,$,
  while the dashed line shows $\left| c_2 \left( t \right) \right|^2$. Since
  there is an odd number of jumps (three jumps in this example), the
  trajectory evolves into the fixed point $\left| c_i \right| = \delta_{i,
  1}$.}
\end{figure}

\subsubsection{Superposition of two localized states}

We consider the expectation value for the coefficients after a jump, that is
\begin{eqnarray}
  \langle c_k \left( q \right) \rangle_G & \assign & \int_{- \infty}^{\infty}
  \mathd q \, G \left( q \right) c_k \left( q \right) \,, \hspace{2em} k = 1,
  2 \, .  \label{eq:Gaverage}
\end{eqnarray}
Upon inserting (\ref{eq:process2}) one obtains $\langle c_1 \left( q \right)
\rangle_G =\mathcal{N}' \left| c_2 \right|^2 c_1$, with the normalization
constant $\mathcal{N}' = \langle \mathcal{N}_q \left( e^{iqx_1 / \hbar} -
e^{iqx_2 / \hbar} \right) \rangle_G \,$, see (\ref{eq:jumpoperator}). Using
$\text{$\left| \langle c_1 \left( q \right) \rangle_G \right|^2 + \left|
\langle c_2 \left( q \right) \rangle_G \right|^2 = 1$}$, we find
$\text{$\left| \mathcal{N}' \right| = 1 / \left( \left| c_1^{} c_2 \right|
\right)$}$ which implies
\begin{eqnarray}
  \left| \langle c_1 \left( q \right) \rangle_G \right| & = & \left| c_2
  \right| \, .  \label{eq:averagejump}
\end{eqnarray}
This shows that after an average jump the moduli of the coefficients are simply
interchanged. This property (which does not hold for $N > 2$) makes the
stochastic process analytically tractable, not least because the dynamics is independent
of the phases of the coefficients. Since the deterministic part
(\ref{eq:coefficients}) of the evolution is monotonic for $N = 2$, a
trajectory starting from $\left| c_1 \left( 0 \right) \right| < 1 / 2$ will
end up in the state $\left| c_i \left( \infty \right) \right| = \delta_{i, 1}$
if and only if an odd number of jumps occurs in the process. This is
demonstrated in Fig.~8. Crucially, the jump rate $r_{\tmop{tot}} \, \left( t
\right)$,
\begin{eqnarray}
  r_{\tmop{tot}} \left( t \right) & \assign & \int_{- \infty}^{\infty} \mathd
  qr_q \left( t \right) \bignone \nonumber\\
  & = & 2 F \left( x_1 - x_2 \right) \left| c_1 \left( t \right) \right|^2
  \left| c_2^{} \left( t \right) \right|^2 \,,  \label{eq:totalrate}
\end{eqnarray}
is unaffected by the jump (\ref{eq:averagejump}) at all times, since it is
invariant under interchanging the coefficients. Hence, the time dependence of
the jump rate is identical for all trajectories, which, in turn, means that
the number of jumps follows an inhomogeneous Poisson process. Therefore, the
probability for an odd number of jumps, which is equal to the statistical
weight $p_1$ of the pointer state $\pi_1 \left( x \right)$, is given by
\begin{eqnarray}
  \tmop{Prob} \left( \tmop{odd} \right) & = & \left( 1 - e^{- 2 \mu \left(
  \infty \right)} / 2 \right) \,,  \label{eq:Probodd}
\end{eqnarray}
with $\mu \left( t \right) = \int_0^t \mathd \tau \, r_{\tmop{tot}} \left(
\tau \right)$ the integrated jump rate. The latter can easily be evaluated by
noting that (\ref{eq:coefficients}) can be written for $N = 2$ as
\begin{eqnarray}
  2 F \left( x_1 - x_2 \right) \left| c_1 \left( \tau \right) \right|^2 \left|
  c_2 \left( \tau \right) \right|^2 & = & \frac{1}{2} \frac{\mathd}{\mathd t}
  \ln \left( 1 - 2 \left| c_1 \left( \tau \right) \right|^2 \right) \, . 
  \label{eq:equation91}
\end{eqnarray}
Upon inserting this result into (\ref{eq:totalrate}) we obtain the integrated
jump rate:
\begin{eqnarray}
  \mu \left( \infty \right) & = & \int_0^{\infty} \mathd \tau \bignone
  \frac{1}{2} \frac{\mathd}{\mathd t} \ln \left( 1 - 2 \left| c_1 \left( \tau
  \right) \right|^2 \right) \, \nonumber\\
  & = & - \ln \left( 1 - 2 \left| c_1 \left( 0 \right) \right|^2 \right) / 2
  \, .  \label{eq:integratedjumprate}
\end{eqnarray}
Noting (\ref{eq:Probodd}) we thus find the probability for an odd number of
jumps
\begin{eqnarray}
  \tmop{Prob} \left( \tmop{odd} \right) & = & \left| c_1 \left( 0 \right)
  \right|^2 \, .  \label{eq:probodd}
\end{eqnarray}
This finally confirms that the statistical weights of the pointer states are
indeed given by the expected overlap (\ref{eq:distribution}).

\begin{figure}[tb]

  \begin{center}
    \resizebox{9cm}{!}{\includegraphics{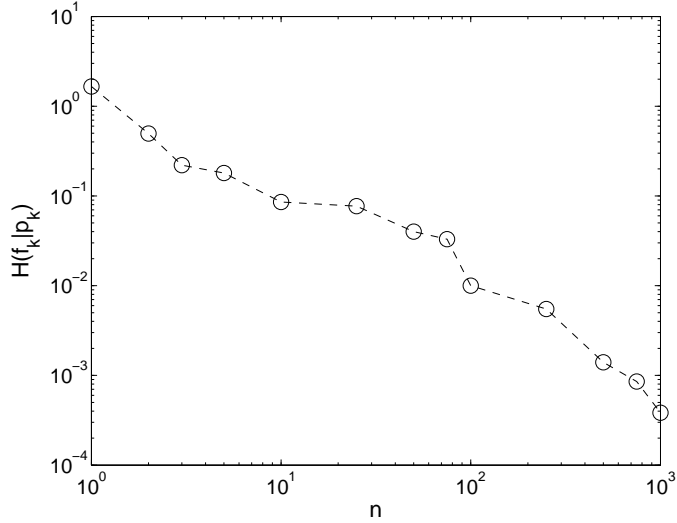}}
  \end{center}
  \caption{Relative entropy $H \left( f_k |p_k \right)$ of the numerically
  obtained distribution of pointer states $f_k$ with respect to the expected
  distribution $p_k = \left| c_k \right|^2$ as a function of the number of
  trajectories $n$ generated in the simulation. The plot indicates that the
  pointer states are distributed according to the initial overlap $\left| c_k
  \right|^2$.}
\end{figure}

\subsubsection{Superposition of $N > 2$ localized states}

The stochastic process is much more involved if the initial superposition
consists of more than two pointer states. Our numerical implementation of the
stochastic process defined by (\ref{eq:coefficients}), (\ref{eq:process2}) and
(\ref{eq:process3}) is based on a Metropolis-Hastings algorithm to draw the
momentum transfer $q$ in accordance with the rate (\ref{eq:process3}), with $G
\left( q \right)$ a Gaussian. Each of the generated trajectories ends
asymptotically in one of the fixed points corresponding to a pointer state,
and we thus obtain a numerical estimate of the statistical weights by means of
the relative frequencies $f_k \:, \, 1 \leqslant k \leqslant N$, of the
asymptotic states. To confirm the expected probability distribution $p_k =
\left| c_k \left( 0 \right) \right|^2$, we evaluate the relative entropy $H
\left( f_k |p_k \right)$ between these two distributions. Figure 9 shows the
result for a random initial state with $N = 5$ as a function of the number of
trajectories $n$, indicating convergence to zero. I{\tmstrong{}}n addition, we
found for $100$ random initial states, with random $2 < N < 11$, based on
$10^4$ trajectories that the relative entropy was always less than $4 \times
10^{- 3}$. This holds both for cases where the initial wave packets $\pi_i
\left( x \right)$ are far apart such that the localization rate is saturated,
$F \left( x_i - x_j \right) \simeq \gamma$, and for situations where the wave
packets are located close together such that $F \left( x_i - x_j \right) <
\gamma$. We conclude that the asymptotic trajectories are indeed distributed
according to the expected overlap (\ref{eq:distribution}). \ \ \ \

As an alternative confirmation of the statistical weights, we performed a
$\chi^2$-test. Similar to the treatment above, $100$ random initial states
$\left\{ \Psi_i | \, 1 \leqslant i \leqslant 100 \right\} \,$, with random $2
< N < 11$, were drawn by the simplex picking method. For each random state, $n
= 100$ trajectories were generated, each of which ends asymptotically in one
of the pointer states. Using the observed relative frequencies $f_k \:, \, 1
\leqslant k \leqslant N$, of the pointer states, we evaluate
\begin{eqnarray}
  \chi^2 & = & n \sum_{k = 1}^N \bignone \frac{\left( f_k - \left| c_k \left(
  0 \right) \right|^2 \right)^2}{\left| c_k \left( 0 \right) \right|^2 } \,, 
  \label{eq:chisquare}
\end{eqnarray}
for each random state. In order to verify that the pointer states are
distributed according to $\left| c_k \left( 0 \right) \right|^2$, the set
$\left\{ \chi_i^2 \right\}$ must be shown to be sampled from a
$\chi_{}^2$-distribution with $N - 1$ degrees of freedom. Comparing the set
$\left\{ \chi_i^2 \right\}$ with the $\alpha$-quantiles
{\tmstrong{}}{\footnote{For instance, the $0.9$ quantile is the value such
that $90\%$ of the samples lie below $Q_{0.9}$.}} (denoted by $Q_{\alpha}$) of
the corresponding $\chi^2$-distribution, a typical run shows ten cases where
$\chi_i^2 > Q_{0.9 \,}$, one case where $\chi_i^2 > \text{$Q_{0.99}$}$, but
not a single case where $\chi_i^2 > \text{$Q_{0.999}$}$, as one expects if the
$\left\{ \chi_i^2 \right\}$ are $\chi^2$-distributed. Like above, this
confirms statistically that the asymptotic trajectories are distributed
according to the expected overlap (\ref{eq:distribution}).

\section{Conclusion}

In this article, we related the nonlinear pure state equation discussed in
{\cite{Diosi2000a,gisin1995a,strunz2002a}} to a specific orthogonal unraveling
of the collisional decoherence master equation. This puts into evidence that
the dynamics of a particle in an ideal gas environment can be represented by
an ensemble of pure state trajectories which evolve into spatially localized
pointer states. For sufficiently strong collisions with the background gas,
these solitonic wave packets move according to the classical equations of
motion, thus explaining the emergence of classical dynamics within the quantum
framework.

Once the pointer state is reached by an individual quantum trajectory, the
latter is no longer affected by the stochastic part of the unraveling, such
that the integration of the trajectory is reduced to the solution of the
classical equations of motion. This suggests that the orthogonal unraveling is
an efficient algorithm for the long time solution of master equations which
exhibit a pointer basis. On the other hand, also the short time solution turns
out to be efficient, since the orthogonal unraveling can be reduced (under
appropriate assumptions) from an infinite dimensional unraveling to a
stochastic process in $\mathbbm{C}^N$.

Future studies might consider the emergence and dynamics of pointer states in
dissipative quantum systems. We note that the present work relies on the model
of pure collisional decoherence which does not describe long time effects such
as dissipation or thermalization. It would certainly be worth to determine the
pointer states of a more involved model such as the quantum linear Boltzmann
equation presented in {\cite{Hornberger2006b,Hornberger2008a,Vacchini2009a}}. For large mass
ratios between the test particle and the gas, one expects that the pointer states then evolve according to a
Langevin equation, thus explaining the emergence of classical Brownian motion
within the quantum framework.

We thank B. Vacchini for helpful discussions. The work was supported by the
DFG Emmy Noether program.

\end{document}